\documentclass{tlp}
\usepackage{aopmath}
\usepackage{graphics}
\usepackage{amssymb,fleqn}
\usepackage{latexsym}
\usepackage{epsfig}

\newtheorem{definition}{Definition} 
\newtheorem{example}{Example} 

\newcommand{\dnext}[1]{\put(3,2){\circle{5}}\;\;\, #1} 
\newcommand*{\dbox}[1]{\put(2,1){\line(1,0){4}}\put(6,1){\line(0,1){4}}\put(6,5){\line(-1,0){4}}\put(2,5){\line(0,-1){4}}\;\;\; (#1)}

\newcommand*{\tccp}{\textsf{tccp}} 
\newcommand*{\cc}{\textsf{cc}} 
\newcommand*{\tcc}{\textsf{tcc}} 
\newcommand*{\ntcc}{\textsf{ntcc}} 
\newcommand*{\hcc}{\textsf{hcc}}

\newcommand*{\anow}{\mathsf{now}} 
\newcommand*{\athen}{\mathsf{then}} 
\newcommand*{\aelse}{\mathsf{else}} 
\newcommand*{\astop}{\mathsf{stop}} 
\newcommand*{\aabort}{\mathsf{abort}} 
\newcommand*{\apar}{||}

\newcommand*{\atell}[1]{\mathsf{tell(}#1\mathsf{)}} 
\newcommand*{\aask}[1]{\mathsf{ask(}#1\mathsf{)}} 
\newcommand*{\aprocedure}[2]{\mathsf{#1(}#2\mathsf{)}}

\newcommand*{\true}{\mathit{true}} 
\newcommand*{\false}{\mathit{false}} 
\newcommand*{\Conf}{\mathbf{Conf}}

\newcommand{\lto}{\longrightarrow} 
\newcommand*{\Var}{\mathcal{V}} 

\newcommand*{\cB}{\mathcal{B}} 
\newcommand*{\cC}{\mathcal{C}} 
 
\newcommand*{\cK}{\mathcal{K}} 
\newcommand*{\cO}{\mathcal{O}} 
\newcommand*{\cQ}{\mathcal{Q}} 
 
\newcommand*{\cU}{\mathcal{U}} 
 
\newcommand*{\ie}   {i.e.,} 
\newcommand*{\wrt}  {w.r.t.}

\title{Automatic Verification of Timed Concurrent Constraint Programs\thanks{This
work has been
partially supported by the EU (FEDER) and the Spanish MEC, under grant
TIN 2004-7943-C04-02, by ICT for EU-India Cross Cultural
Dissemination Project under grant ALA/95/23/2003/077-054, and
by the Italian project Cofin'04
AIDA.}}
\author[M. Falaschi and A. Villanueva]{Moreno Falaschi\\
Dip. Matematica e Informatica, University of Udine\\
         Via delle Scienze, 206. I-33100 Udine, Italy\\
E-mail: {\tt falaschi@dimi.uniud.it}\vspace{1ex}\\
\protect{\normalsize\emph{Alicia Villanueva}}\\
Dep. Sistemas Inform\'aticos y Computaci\'on, Technical University of Valencia\\
         Camino de Vera s/n. E-46022 Valencia, Spain\\
E-mail: {\tt villanue@dsic.upv.es}
}

\submitted{29 December 2003}
  \revised{7 April 2004}
  \accepted{10 May 2005}

\shorttitle{ Automatic Verification of Timed Concurrent Constraint Programs}

\begin{document}

\maketitle

\bibliographystyle{acmtrans}


\begin{abstract}
The language \emph{Timed Concurrent Constraint} (\tccp{}) is the
extension over time of the \emph{Concurrent Constraint
Programming} (\cc{}) paradigm that allows us to specify
concurrent systems where timing is critical, for example
\emph{reactive systems}. Systems which may have an infinite number
of states can be specified in \tccp{}. \emph{Model checking} is
a technique which is able to verify finite-state systems with a
huge number of states in an automatic way. In the last years
several studies have investigated how to extend model
checking techniques to systems with an infinite number of states.
In this paper we propose an approach which exploits the
computation model of \tccp{}. Constraint based computations
allow us to define a methodology for applying a model checking
algorithm to (a class of) infinite-state systems. We extend the
classical algorithm of model checking for LTL to a specific logic
defined for the verification of \tccp{} and to the \tccp{}
Structure which we define in this work for modeling the program
behavior. We define a restriction on the time in order to get a
finite model and then we develop some illustrative examples. To
the best of our knowledge this is the first approach that defines
a model checking methodology for \tccp{}.
\end{abstract}

\begin{keywords}
Automatic verification, reactive systems, timed concurrent
constraint programming, model checking
\end{keywords}

\section{Introduction}\label{introduction}

\emph{Model checking} is a technique for formal verification that was
defined for finite-state systems. It was first introduced in
\cite{CE81} and \cite{QS82} for verifying \emph{automatically} if
a system satisfies a given property. Concurrent systems can be very
complicated, and the process of modeling and verifying them by hand
can be hard. Thus, the development of formal and fully automatic
methods such as model checking is essential. Basically, this technique
consists in an exhaustive analysis of the state-space of the system.
This exhaustive analysis implies that, in principle, we can apply it
only to finite-state systems limiting a lot its applicability.
Furthermore, the state-explosion problem is the main drawback even for
finite-state systems and for this reason many approaches in the
literature try to mitigate it.  Two of the main solutions for the
state-explosion problem that have been presented in the last years are
the symbolic approach \cite{McM93} and the algorithms for abstract model checking
\cite{Dam96}.  The idea which is shared by these approaches
is to reduce the number of states of the system.

The different approaches to the model checking problem for infinite
state systems can be classified in two categories. The first one
corresponds to those approaches that construct an abstract finite
model of the system which can be automatically verified (see
\cite{CGL94,LGS+95}). The second category contains those approaches
based on the symbolic reachability analysis where a finite
representation of the set of reachable configurations of the system is
calculated (see \cite{ACH+95,CH78,BEM97,BG96}).  The methodologies
that make use of regular languages and regular relations are
considered in the so called \emph{regular model checking} approach
\cite{PS00,KMM+97,BJNT00}. Moreover, in \cite{AAB+99} the notion of
abstraction and the notion of symbolic reachability are combined in
order to define a method to verify infinite-state systems. Our
approach is novel and makes use of a notion of abstraction based on
constraints and a time interval. The notion of constraints is used to
collapse the number of states.

In \cite{MP95} \emph{reactive systems} are defined as those systems
that keep exchanging information with their environment at run time.
Reactive systems are typically defined as a set of processes working
in parallel, hence the family of reactive systems is strictly related
to the notion of \emph{concurrency}.  In some cases it is not expected
that the system terminates but it may continue its execution
indefinitely. Examples of such systems are operating systems,
communication protocols or some kind of embedded systems. Thus it is
quite useful to have a specification language that supports
concurrency which makes easier for the user to describe systems.
Usually, in model checking, by exploiting concurrency we model the
whole system, including the environment. For example, users are
represented as a concurrent process which models the possible actions
that users can perform to interact with the system.

The language \emph{Temporal Concurrent Constraint Programming}
(\tccp{}) extends the \emph{Concurrent Constraint Programming} (\cc{})
paradigm defined in \cite{Sar89} with a notion of time. This extension
is suitable for modeling reactive systems.  Actually, in the
literature you can find two similar languages which extend \cc{} with
some notion of time: the \tcc{} language first presented in
\cite{SJG94a} and the \ntcc{} language defined in \cite{NPV02a}.
\tccp{} is a declarative language defined in \cite{BGM99} that handles
constraints which is a key characteristic for the results which we
achieve in the present work. Our idea is to take advantage of the
natural properties of the language in order to define a model-checking
algorithm that allows us to verify reactive systems specified in
\tccp{}. Note that when we speak of reactive systems we are not
limiting ourselves to finite-state systems. The \tccp{} language
allows us to model infinite-state systems, hence we tackle the problem
of model checking for infinite-state systems. We show how the
constraint nature of the language and the fact that it has a built-in
notion of time can be exploited usefully.

Some related works can be found in the literature where constraints
are used for solving similar problems. In \cite{DP99,DP01} the authors
present a method that allows them to verify a communication protocol
with an infinite number of states in the sense that they prove that a
client-server protocol is correct for an arbitrary number of processes
(clients). This could not be proved by using classical approaches to
model checking, however it become possible thanks to the use of the
notion of constraint.

The model-checking technique can be divided into three main phases;
specification, modeling and verification. In this work, we use the
notion of constraint in the three phases of the model-checking
technique. First, we introduce the notion of constraint in the
constructed model of the system. We note that constraints are able to
represent in a compact manner a set of possible values that the system
variables can take (i.e., a possibly infinite set of states if we use
the classical notion of state). In the second phase we use a logic
able to handle constraints for specifying the property to be verified.
Such logic was presented in \cite{BGM01} and revisited in
\cite{BGM02}.  The last phase of the model-checking technique consists
in defining an algorithm that determines whether the system satisfies
the property by using the two outputs of the previous phases. In this
work we extend the classical algorithm defined for LTL to the
constrained approach. Note that we can take as a reference the
classical algorithm because we use a logic able to handle constraints,
and this makes possible to combine it with the \tccp{} Structure
defined in this paper to model the system. Since this structure
contains constraints, it would not be possible to use a classical
temporal logic directly. To the best of our knowledge this is the
first time that a model-checking algorithm for systems specified with
the \tccp{} language is defined.  Some of the results in this work
have been included in Villanueva's doctoral thesis \cite{Vil03}.

In \cite{FPV00a,FPV00b} we presented a framework that allowed us to
build a graph structure as a first step for applying the
model-checking technique to \tcc{} programs. \tcc{} is a language
similar to \tccp{} for programming embedded systems. The main
differences between \tcc{} and the language that we consider here is
in the deterministic nature of the \tcc{} language versus the
non-determinism, and the \emph{monotonicity} of the store in
\tccp{}. Monotonicity means that the store of the system always
increases. \tcc{} is not monotonic since the store is reset when
passing from one time instant to the following one.  These differences
make the graph structures defined in \cite{FPV00a,FPV00b} and in this
work completely different. We will show these differences in detail in
the following sections.  Moreover, only the modeling process of the
method was presented in \cite{FPV00a,FPV00b}, whereas in this paper we
provide the logic used for the specification of the property and the
model-checking algorithm as well.

This paper is organized as follows. In Section~\ref{sec:preliminaries}
we introduce some basic theoretic notions. In
Section~\ref{sec:tccp-language} we present the basic notions of the
\tccp{} language. Then, in Section~\ref{sec:modelling} we describe the
method to construct an adequate model of the system, which is shown to
model correctly the language operational semantics. In
Section~\ref{sc:the-logic} we present the logic for specifying the
properties of our system.  In Section~\ref{sec:the-algorithm} we
define the algorithm that applies the model-checking technique to this
model and show its correctness.  Section~\ref{related-work} discusses
some related work.  Finally, in Section~\ref{conclusions} final
remarks and future work are discussed.

\section{Preliminaries}\label{sec:preliminaries}

In this section we present some definitions necessary to follow the
technical details of this work. For a quick reading it is possible to
skip to Section~\ref{sec:tccp-language}.

A \emph{Constraint System} is a system of partial information. We
follow the definition of Saraswat \emph{et al.}:

\begin{definition}[Simple constraint system \cite{SRP91}]\label{def:constraint-system}
  Let $D$ be a non-empty set of \emph{tokens} or \emph{primitive
    constraints}. A \emph{simple constraint system} is a structure $\langle
  C,\vdash\rangle$ where $\vdash \subseteq \wp_f(C) \times C$ is an
  \emph{entailment relation} satisfying:
\begin{description}
\item[C1] $u\vdash P$ whenever $P\in u$,
\item[C2] $u\vdash Q$ whenever $u\vdash P$ for all $P\in v$ and $v\vdash Q$.
\end{description}
\end{definition}

Moreover, an element of $\wp_f(C)$ is called a \emph{finite
  constraint} and $\vdash$ is extended to $\wp_f(C)\times \wp_f(C)$ in
the obvious way. Finally, $u\approx v$ iff $u\vdash v$ and
$v\vdash u$. We also say that $u\geq v$ when $v\vdash u$.

\begin{definition}[Cylindric constraint system
  \cite{SRP91}]\label{def:cylindric-constraint-system} We define a
  cylindric constraint system as a structure $\langle C, \vdash,
  \mathcal V, \{ \exists_x \mid x \in \mathcal V \} \rangle$ such that
  $\langle C, \vdash \rangle$ is a simple constraint system, $\mathcal
  V$ is an infinite set of \emph{variables} and, for each $x \in
  \mathcal V$, $\exists_x: \wp_f(C) \rightarrow \wp_f(C)$ is an
  operation satisfying:
  \begin{description}
  \item[E1] $u \vdash \exists_x u$,
  \item[E2] $u \vdash v$ implies $\exists_x u \vdash \exists_x v$,
  \item[E3] $\exists_x (u \sqcup \exists_x v) \approx \exists_x u \sqcup
    \exists_x v$,
  \item[E4] $\exists_x \exists_y u \approx \exists_y \exists_x u$.
  \end{description}
  $\exists_x$ is called the \emph{existential quantifier} or
  \emph{cylindrification operator}.

\smallskip

A set of \emph{diagonal elements} for a cylindric constraint system is
a family $\{\delta_{xy} \in C \mid x,y \in \mathcal V \}$ such that
  \begin{description}
  \item[D1] $\emptyset \vdash \delta_{xx}$,
  \item[D2] if $y \neq x,z$ then $\{ \delta_{xz} \} \approx \exists_y
    \{\delta_{xy}, \delta_{yz}\}$,
  \item[D3] if $x \neq y$ then $\{\delta_{xy}\} \sqcup \exists_x (u \sqcup
    \{\delta_{xy}\}) \vdash u$.
  \end{description}
\end{definition}

We define an \emph{element} $c$ of a cylindric constraint
system $\langle C, \vdash \rangle$ as a subset of $C$ closed by
entailment, \ie{} such that $u \subseteq_f c$ and $u \vdash P$ implies
$P \in c$.

\section{Timed Concurrent Constraint Language}\label{sec:tccp-language}

The \tccp{} language was developed in \cite{BGM99}. It was designed as
a computational model which allows one to model reactive and real-time
systems. Thus, it is possible to specify and to verify distributed,
concurrent systems where the notion of time is a crucial question.
\tccp\ is based on the \cc\ paradigm \cite{Sar89,SR90,SRP91} that was
presented as a general concurrent computational model.

The computational model of \cc{} is defined by means of
a global store and a set of defined agents that can add
(tell) information into the store or check (ask) whether a constraint
is entailed by the store. Computations evolve as an accumulation of
information into a global store. In \tccp{} the agents defined for
\cc{} are inherited. The model is enriched with a new agent and a
\emph{discrete global clock}. It is assumed that ask and tell actions
take one time-unit and the parallel operator is interpreted in terms
of maximal parallelism. Computation evolves in steps of one time-unit.
It is assumed that the  response time of the constraint solver is
constant, independently of the size of the store.  In practice
some restrictions (mentioned below) are taken in order to ensure that
these hypothesis are reasonable (the reader can see \cite{BGM99} for
details).

To model reactive systems it is necessary to have the ability to
describe notions as \emph{timeout} or \emph{preemption}.  The
timeout behavior can be defined as the ability to wait for a specific
signal and, if a limit of time is reached and such signal is not
present, then an exception program is executed. The notion of
preemption is the ability to abort a process when a specific signal is
detected.  In \tccp{} these behaviors can be modeled by using the new
conditional agent (not present in \cc)
$$\anow\, c\, \athen\, A\, \aelse\, B$$
which tests if in the current
time instant, the store entails the constraint $c$ and if it occurs,
then in the same time instant it executes the agent $A$; otherwise, it
executes $B$ (in the same time instant).  A limit for the number of
nested conditional agents is imposed in order to ensure the bounded
time response of the constraint solver within a time instant.

\subsection{Syntax}\label{sec:syntax}

The \tccp\ language is parametric to an underlying cylindric
constraint system as defined in Section~\ref{sec:preliminaries}.
Since now we assume that $\cC = \langle
C,\vdash,\Var,\exists\rangle$ is the underlying constraint system
for \tccp{}.  Given $\cC$, in Figure~\ref{fig:tccp-syntax} we show
the syntax of the agents of the language. We assume that $c$ and
$c_i$ are finite constraints (i.e. elements) in $\cC$.

\begin{figure}[ht]
    \centering {\fbox{
    \begin{minipage}{21cm}
        \begin{tabbing}
            (Agents) \hspace{2cm} \= A  \= ::= \= $\atell{c}$ \hspace{2.5cm} \= --
            Tell \\
            \>    \> $\mid$  \>  $\astop$  \> -- Stop \\
            \>    \> $\mid$  \>  $\sum_{i=1}^{n}\aask{c_i}\rightarrow
            A_i$ \> -- Choice \\
            \>    \> $\mid$  \>  $\anow\, c\, \athen\, A\, \aelse\, A$ \> --
            Conditional \\
            \>    \> $\mid$  \>  $A\mid\mid A$ \> -- Parallel \\
            \>    \> $\mid$  \>  $\exists x\, A$ \> -- Hiding \\
            \>    \> $\mid$  \>  $\aprocedure{p}{x}$ \> -- Procedure Call \\
            (Declarations)       \> D \> ::= \> $D\ldotp D$\\
            \>    \> $\mid$  \> $\aprocedure{p}{x} $:-$ A$\\
            (Program)            \> P \> ::= \> $D\ldotp A$
        \end{tabbing}
    \end{minipage}
    }}\caption{\tccp{} syntax (following F. de Boer \emph{et al.})}%
\label{fig:tccp-syntax}
\end{figure}

The Parallel and Hiding agents are inherited from the \cc\ model
and behave in the same way. Thus, the Parallel agent represents
concurrency, whereas the Hiding operator makes a variable local to
some process. Also the Tell, Choice and Procedure Call agents were
present in the \cc{} model, but in \tccp{} they have a different
semantics since in the timed model, these three agents cause
extension over time. The Tell agent adds the information $c$ to
the store, but this information is available to other agents only
in the following time instant.  Therefore, we can say that the
tell action takes one unit of time. The same thing occurs with the
Choice and Procedure Call agents.  Thus, when we execute the
$\sum_{i=1}^{n}\aask{c_i}\rightarrow A_i$ agent, the execution of
$A_i$ starts in the next time instant.  Note that the Choice agent
models the nondeterministic behavior of the language, thus
nondeterminism is always associated to a time delay.

Finally, the Conditional agent ($\anow\,c\,\athen\,A\,\aelse\,B$) is
the new agent introduced in the model in order to capture negative
information. It behaves within a time unit in the sense that the
condition is checked \emph{in the same instant of time} as the
execution of the corresponding agent is started. In particular, if the
guard is satisfied, then $A$ will be executed, otherwise the agent $B$
will be executed (we note that $B$ is executed also in the case when
the store entails neither $c$ nor $\neg c$). If we have two nested
conditional agents, then the guards are recursively checked within the
same time instant. This is the reason why \tccp{} needs a
restriction about the maximum number of nested conditional agents.


\subsection{\tccp{} Operational Semantics}\label{sec:operational-semantics}

In Figure~\ref{fig:tccp-semantics} it is shown the operational
semantics for \tccp{} as described in \cite{BGM99}. Each transition
step takes one unit of time. In a configuration ($\Conf$) there are
two components: a set of agents and a finite constraint representing
the store. The transition relation
$\longrightarrow\subseteq\Conf\times\Conf$ is the least relation that
satisfies the rules in Figure~\ref{fig:tccp-semantics}.  We can say
that the transition relation characterizes the (temporal) evolution of
the system.


\begin{figure}[ht]
    \centering{\fbox{
    \begin{minipage}{21cm}
        \begin{tabbing}
            \textbf{R1}\hspace{.5cm}\= $\langle\atell{c},d\rangle
            \longrightarrow \langle\astop,c\sqcup
            d\rangle$\hspace{2cm}\=\\[1.3em]
            \textbf{R2}\> $\langle\sum_{i=1}^{n}\aask{c_i}
            \rightarrow A_i,d\rangle\longrightarrow\langle A_j,d\rangle$\>
            $j\in [1,n]\; \text{and}\; d\vdash c_j$\\[1.3em]
            \textbf{R3}\> {\Large $\frac{\langle A,d\rangle\longrightarrow\langle
            A',d'\rangle} {\langle\anow\,c\,\athen\,A\,\aelse
            \,B,d\rangle\longrightarrow
            \langle A',d'\rangle}$} \> $d\vdash c$\\[1.3em]
            \textbf{R4}\> {\Large $\frac{\langle A,d\rangle \not\longrightarrow}
            {\langle\anow\,c\,\athen\,A\,\aelse
            \,B,d\rangle\longrightarrow
            \langle A,d\rangle}$} \> $d\vdash c$\\[1.3em]
            \textbf{R5}\> {\Large $\frac{\langle B,d\rangle\longrightarrow\langle
            B',d'\rangle} {\langle\anow\,c\,\athen\,A\,\aelse
            \,B,d\rangle\longrightarrow
            \langle B',d'\rangle}$}\> $d\not\vdash c$\\[1.3em]
            \textbf{R6}\> {\Large $\frac{\langle B,d\rangle \not\longrightarrow}
            {\langle\anow\,c\,\athen\,A\,\aelse
            \,B,d\rangle\longrightarrow
            \langle B,d\rangle}$} \> $d\not\vdash c$\\[1.3em]
            \textbf{R7}\> {\Large $\frac{\langle A,c\rangle \longrightarrow \langle
            A',c'\rangle\;\; \langle B,c\rangle \longrightarrow\langle B',d'\rangle}
            {\langle A\mid\mid B,c\rangle\longrightarrow\langle A'\mid\mid
            B',c'\sqcup d'\rangle}$}\\[1.3em]
            \textbf{R8}\> {\Large $\frac{\langle
            A,c\rangle\longrightarrow\langle A',c'\rangle\;\; \langle
            B,c\rangle\not\longrightarrow}{\langle A\mid\mid B,c\rangle\longrightarrow
        \langle A'\mid\mid B,c'\rangle}$} \\[1.3em]
    \textbf{R9}\> {\Large $\frac{\langle
            A,c\rangle\longrightarrow\langle A',c'\rangle\;\; \langle
            B,c\rangle\not\longrightarrow}{\langle B\mid\mid A,c\rangle\longrightarrow
        \langle B\mid\mid A',c'\rangle}$} \\[1.3em]
    \textbf{R10}\> {\Large $\frac{\langle A,d\sqcup \exists_x
    c\rangle\longrightarrow \langle B,d'\rangle} {\langle\exists^d
    xA,c\rangle\longrightarrow
    \langle\exists^{d'}xB,c\sqcup\exists_x d'\rangle}$}\\[1.3em]
    \textbf{R11}\> $\langle \mathsf{p}(x),c\rangle\longrightarrow\langle A,c\rangle$ \>
    $\mathsf{p}(x) :- A\in D$
\end{tabbing}
\end{minipage}
}} \caption{Operational semantics for \tccp{} language extracted from F. de Boer~\emph{et al.}}
\label{fig:tccp-semantics}
\end{figure}

Since \tccp{} interprets concurrency in terms of \emph{maximal
  parallelism}, we assume that there are as many processors as
needed to execute a program. This behavior is described by means of
rules \textbf{R7}, \textbf{R8} and \textbf{R9} where the reader can
see that whenever it is possible, two agents are executed concurrently.

Rules \textbf{R3}, \textbf{R4}, \textbf{R5} and \textbf{R6} describe
the operational semantics for the conditional agent. Note that the
different possible behaviors depend on the store and on the initial
configuration.  Rule \textbf{R10} shows the semantics for the Hiding
operator.  Intuitively, the rule says that, if there exists a
transition $\langle A,d\sqcup \exists_x c\rangle\longrightarrow
\langle B,d'\rangle$, then $d'$ is the local information produced by
$A$; moreover, this local information $d'$ must be hidden from the
main process.

 The observable behavior of the language is defined from the
 transition system described in Figure~\ref{fig:tccp-semantics} and
 considers the input/output of finite and infinite computations:

 \begin{definition}[Observable]\label{def:observable}
 Let $A$ be an agent from the \tccp{} language, the
 operational behavior is given by the set of resulting stores computed by
 $A$ for each given input store, considering finite and infinite
 computations.
 $$\cO(A) = \{d \mid\langle A,c\rangle\lto \ldots\langle
 B,c_{i}\rangle\lto \ldots, \mbox{ where }
 d \equiv  \{ c, c_1, ..., c_i, ... \}\}$$
 \end{definition}

\subsection{Practical Example}

\begin{figure}[ht]
    \begin{center}
    \fbox{
       \scalebox{0.4}{\includegraphics{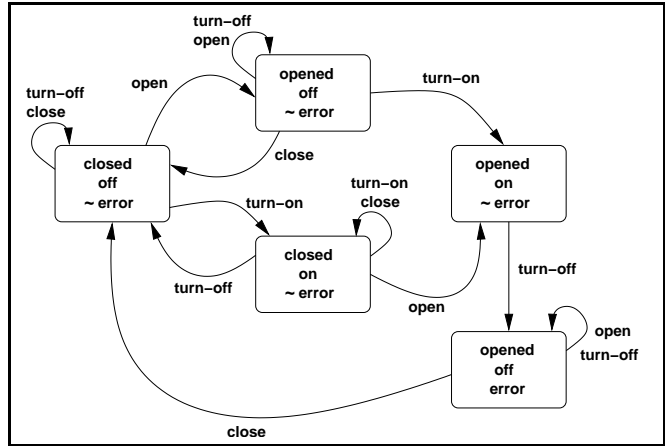}}
    }
    \end{center}
    \caption{Example: the microwave system.} \label{fig:microwave}
\end{figure}

We can find in the literature a variety of examples of systems that
can be modeled using the \tccp{} language. Here we develop a typical
system: a microwave oven. In Figure~\ref{fig:microwave} the reader can
see the behavior of a microwave. We can note that, for example, if we
are in a state where the door of the microwave is closed, the system
is turned-off and no error is detected. If, from that state, we open
the door, then we move to the state on the top of the figure.

The whole system example is inspired in the system for a microwave
control shown in the classical literature \cite{CGP99}. However, we
have considered only a subpart of the system in order to easily use
this example as a reference in this work.
In Figure~\ref{fig:tccp-example} we show the \tccp{} program which
models a reduced part of the microwave system. In particular, it
models the part of the system which detects if the door is open when
the microwave is turned-on.

\begin{figure}[ht]
    \centering{ \fbox{
    \begin{minipage}{21cm}
        \begin{tabbing}
            \quad \= \quad \= \quad \= \quad \=\quad \= \quad \=\quad \= \quad
            \=\quad \= \quad \=\quad \= \quad \=\kill
            $\aprocedure{microwave\_error}{\mathrm{Door},\mathrm{Button},\mathrm{Error}}$ :- \\
            \>\>\> $\exists\,D,B,E($ $\atell{\mathrm{Error} = [\_|E]}$ $||$ $(\atell{\mathrm{Door} = [\_|D]}$ $||$ $(\atell{\mathrm{Button} = [\_|B]}$ $||$\\
            \>\>\>\>\> $(\anow\, (\mathrm{Door}=[\mathrm{open}\mid D]
                        \wedge \mathrm{Button}=[\mathrm{on}\mid B])\, \athen$\\
            \>\>\>\>\>\>\>\> $(\exists\, E1(\atell{\mathrm{E}=[\mathrm{yes}\mid E1]}) \apar$\\
            \>\>\>\>\>\>\>\> $\exists\, B1(\atell{B=[\mathrm{off}\mid B1]}))$\\
            \>\>\>\>\> $\aelse$\\
            \>\>\>\>\>\>\>\> $\exists\, E1(\atell{\mathrm{E}=[\mathrm{no}\mid E1]}) \apar$\\
            \>\>\>\>\> $\aprocedure{microwave\_error}{D,B,E})))).$
        \end{tabbing}
    \end{minipage}
    }} \caption{Example of a \protect{\tccp{}} program: a simple
    error controller}\label{fig:tccp-example}
\end{figure}

Looking into the program code, we can observe that a Conditional agent
checks if the door is open when the microwave is turned-on. In that
case, it forces (with the Tell agent) that in the following time
instant, the microwave is turned-off and an error signal is emitted.
If it is not true that the door is open and the microwave is working
on, then the program simply emits (via the Tell agent) a signal of
\emph{no} error that will be available in the global store in the
following time instant.  Therefore, this example corresponds to the
part of the system which avoids wrong behaviors such as those in
Figure~\ref{fig:microwave} which are represented by the two states on
the right.

This simple example allows us to illustrate the fact that \tccp{} is
not able to model strong preemption, i.e., it is not possible to
turn-off the start button in the same instant when the error is
detected. Actually, it is possible to start the execution of the agent
that turns the button off, but the fact that it has been turned off is
visible only in the following time instant.

\section{\tccp{} Model Checking}\label{sec:modelling}

The \cc{} paradigm has some interesting features which allow us to
define a model-checking algorithm for reactive systems. We define a
model-checking algorithm which uses a time interval (provided by the
user) in order to restrict the state-space of the system in the cases
when the algorithm does not terminates. The fact that the time is in
the semantics makes reasonable the use of such restriction since the
user knows how much time is needed to have a response from the system.
The reader could think that the restriction to an interval of time
could make the algorithm incomplete in too many cases. In the
following sections we show that the time interval is not always used.
Obviously, the user must provide a reasonable time interval.
Moreover, if the limit is reached and the verification is terminated,
then we obtain an over-approximation of the system thus some
properties can still be checked. The idea to limit the verification to
a time limit is not new. It has been used in different approaches,
for example in \cite{AHW97}.

Let us now develop a model-checking technique to \tccp{} programs. The
key ideas are that we use the notion of constraint which is underlying
the language in order to have a compact model of the system first, and
second, to handle the model directly to verify properties.

In the following we describe in detail the three main phases which
implement the model-checking algorithm. We also illustrate each phase
with the application of the method to the microwave example.

\subsection{Model Construction}

The first task of the method corresponds to the construction of the
model of the system. In classical approaches, \emph{Kripke
  Structures}\footnote{Kripke Structures were defined in \cite{HC68}.
  The definition can also be seen in \cite{CGP99} for example.} are
used to model the system behavior; in our approach we define a
similar structure called \tccp{} Structure whose states are
essentially a conjunction of constraints of the underlying constraint
system. The idea is to automatize the construction of the model of the
system from the specification. In other words, we take a program
written in \tccp{}, and a model of the system behavior is constructed
in an automatic way.

\subsubsection{Program Labeling}
First of all, we need a labeled version of the specification in order
to construct the model of the system automatically. We adapt the idea
introduced in \cite{MP95} to our framework: a different label is
assigned to each occurrence of an agent. Labels allow us to identify
during the model construction in which point of the execution of the
program we are. The presence or absence of a label determines if the
associated agent can be executed or not during the computation. The
labeling process consists on the introduction of a different label
for each occurrence of a language construct:
\begin{definition}\label{tccp-labelling}
  Let $P$ be a specification, the labeled version $P_l$ of $P$ is
  defined as follows. The subindex $k\in \mathbb{N}$ corresponds to
  the number of labels introduced up to a given point. When the
  labeling process starts, $k=0$ and each time that we introduce a
  new fresh label, $k$ is incremented by one.
    \begin{itemize}
        \item If $P = \astop$ then $P_l = l_{\astop_k} \astop$.
        \item If $P = \atell{c}$ then $P_l = l_{\mathsf{tell}_k} \atell{c}$.
        \item If $P = \sum_{i=1}^{n}\aask{c_i}\rightarrow
            A_i$ then $P_l = l_{\mathsf{ask}_k} \sum_{i=1}^{n} \aask{c_i}\rightarrow  A_l$.
        \item If $P = \anow\, c\, \athen\, A\,\aelse\, B$ then
        $P_l = l_{\anow_k} \anow\, c\, \athen\, A_l\, \aelse\, B_l$.
        \item If $P = A\apar B$ then $P_l = l_{\apar_k} (A_l\apar B_l)$.
        \item If $P = \exists x\,A$ then $P_l = l_{e_k} \exists x\,A_l$.
        \item If $P = \aprocedure{p}{x}$ then $P_l = l_{\mathsf{p}_k} \aprocedure{p}{x}$.
    \end{itemize}
    The labeling of a declaration $D$ of the form $\aprocedure{p}{x}$
    :- $A$ is defined as $l_{\mathrm{p}_i} \aprocedure{p}{x}$ :- $A_l$,
    called $D_l$.  Finally, the labeled version of a program of the
    form $D\ldotp A$ is $D_l\ldotp A_l$.
\end{definition}
In practice, we explore the \tccp\ specification, and each time that
we find an occurrence of a construct we introduce a new label which
identifies such point of the program.

In Figure~\ref{fig:tccp-labelled-example} we show the labeled version
of the microwave error detection program showed in
Figure~\ref{fig:tccp-example}. Note that the structure of the program
has not changed, simply some labels have been added.

\begin{figure}[ht]
    \centering{ \fbox{
    \begin{minipage}{21cm}
        \begin{tabbing}
            \quad \= \quad \= \quad \= \quad \=\quad \= \quad \=\quad \= \quad
            \=\quad \= \quad \=\quad \= \quad \=\kill
            $\{l_{\mathrm{p}_0}\}\,\aprocedure{microwave\_error}{\mathrm{Door},\mathrm{Button},\mathrm{Error}} :-$ \\
            \>\>\> $\{l_{\mathrm{e_1}}\}\,\exists\,D,B,E( \{l_{||_2} (\{l_{t_3}\} \atell{\mathrm{Error} = [\_|E]} ||$ $(\{l_{\mathrm{||_4}}\} (\{l_{\mathrm{t_5}}\}\,\atell{\mathrm{Door} = [\_|D]}$ $||$ \\
            \>\>\>\>\hspace{2.5cm}$\{l_{\mathrm{||_6}}\}(\{l_{t_7}\} \atell{\mathrm{Button} = [\_|B]}$ $||$\\
            \>\>\>\>\> $\{l_{\apar_8}\}\, (\{l_{\mathrm{now}_9}\}\, \anow\, (\mathrm{Door}=[\mathrm{open}\mid D]
                        \wedge \mathrm{Button}=[\mathrm{on}\mid B])\, \athen$\\
            \>\>\>\>\>\>\>\> $\{l_{\apar_{10}}\}\, (\{l_{\mathrm{e_{11}}}\}\, \exists\, E1(\{l_{\mathrm{t_{12}}}\}\, \atell{\mathrm{E}=[\mathrm{yes}\mid E1]}) \apar$\\
            \>\>\>\>\>\>\>\> $\{l_{\mathrm{e_{13}}}\}\, \exists\, B1(\{l_{\mathrm{t_{14}}}\}\, \atell{B=[\mathrm{off}\mid B1]}))$\\
            \>\>\>\>\> $\aelse$\\
            \>\>\>\>\>\>\>\> $\{l_{\mathrm{e_{15}}}\}\, \exists\, E1(\{l_{\mathrm{t_{16}}}\}\, \atell{E=[\mathrm{no}\mid E1]}) \apar$\\
            \>\>\>\>\> $\{l_{\mathrm{p}_{17}}\}\,
            \aprocedure{microwave\_error}{D,B,E})))).$
        \end{tabbing}
    \end{minipage}
    }} \caption{Example of a labeled \protect{\tccp{}} program: a simple
    error controller}\label{fig:tccp-labelled-example}
\end{figure}

\subsubsection{The \protect{\tccp} Structure}

The main point in the modeling phase is the construction of the
graph structure which
represents the system behavior. We define a new graph structure to
represent the system. The \tccp{} \emph{Structure} is defined as a
variant of the \emph{Kripke Structure}. Intuitively, a Kripke
Structure is a \emph{finite} graph structure where there could be many
initial nodes and each node is always related to another one (or to
itself).  Moreover, each state has associated a set of atomic
propositions which are true in such state.

The main difference between the two structures is that the definition
of a state in the Kripke Structure follows the classical notion of
state whereas in our structure, a state consists of a conjunction of
constraints and intuitively it can be seen as a set of classical
states.

Let us formally define our \tccp{} Structure.

\begin{definition}
  The set $AP$ of atomic propositions is defined as the set of
  elements\footnote{See the definition in
    Section~\ref{sec:preliminaries}.}  of the cylindric constraint
  system $\cC$ of the \tccp{} language.
\end{definition}

In the rest of the paper we abuse of notation by identifying the
meaning of the terms \emph{constraint}, \emph{atomic proposition} and
\emph{element}. Next we define what a state of the \tccp{} Structure
is:
\begin{definition}[\protect{\tccp} State]\label{def:tccp-state}
  Let $AP$ be the atomic propositions in the \tccp{} syntax and $L$ be
  the set of all labels generated during the labeling process
  described above. We define the set of states as $S\subseteq
  2^{AP}\times 2^L\;$.
\end{definition}

Before the definition of the \tccp{} Structure, we define the
notion of equivalent states. For this, we need the classical
notion of renaming of variables. Let $y_{1}, \ldots, y_{n}$ be $n$
distinct variables. The substitution $\{x_{1}/y_{1}, \ldots,
x_{n}/y_{n}\}$ is a {\em renaming}.

\begin{definition}[Equivalent States]\label{def:eq-states}
Given two \tccp{} states $s$ and $s'$, we say that the two states are
equivalent if:
\begin{itemize}
\item the set of labels $l\subseteq L$ of $s$ and the set of
labels $l'\subseteq L$ coincide and,
\item there exists a renaming $\gamma$ of variables of the
constraints in $s$ which makes them syntactically identical to the set
of constraints of $s'$
\end{itemize}
\end{definition}

In Definition~\ref{def:tccp-structure}, we define the \tccp{}
Structure. Observe that the differences \wrt{} a Kripke Structure
are the definition of state (in Definition~\ref{def:tccp-state})
and the two labeling functions $C$ and $T$ which replace the
labeling function $L$ of the classical Kripke Structure.

\begin{definition}[\tccp{} Structure]\label{def:tccp-structure}
  Let $AP$ be a set of atomic propositions, we define a \tccp{}
  Structure $M$ over $AP$ as a five tuple $M = (S,S_0,R,C,T)$ where
    \begin{enumerate}
        \item $S$ is a finite set of states.
        \item $S_0\subseteq S$ is the set of initial states.
        \item $R\subseteq S\times S$ is a transition
          relation.
        \item $C : S\rightarrow 2^{AP}$ is the function that returns
        the set of atomic propositions in a given state.
        \item $T : S\rightarrow 2^{L}$ as the function that returns
        the set of labels in a given state.
    \end{enumerate}
\end{definition}

We assume that a transition in the graph represents an increment
of one time-unit in the system. Intuitively, $C$ labels a state
with the set of constraints true in such state. In other words,
this function represents the new information that we know in a
specific instant. $T$ labels each state with the set of labels
associated to agents that must be executed in the following time
instant. In other words, $T$ represents the point of execution in
each instant (or state).

When two states $s$ and $s'$ are related by $R(s,s')$, it means
that it is possible to reach the state $s'$ from state $s$ by
executing the agents associated to the labels in $T(s)$ with the
store $C(s)$ deriving as a result (by applying the renaming
$\gamma$) the store $C(s')$ and the point of execution $T(s')$. In
other words, given a state $s$ and a renaming $\gamma$, we
obtain a state $s'$ whose store is $C(s')\cdot\gamma$.

Given a \tccp{} Structure $Z=(S,S_0,R,C,T)$, we define
$\mathop{tr(Z)}$ as the set of sequences of states of $Z$ starting
from an initial state and which are related by $R$:

\begin{equation}
\mathop{tr}(Z) = \{s\mid s = s_0\cdot s_1\cdots s_n\cdots\; \wedge
s_0\in S_0 \wedge \forall i\geq 0, \exists R(s_i,s_{i+1})\}
\label{eq:traces}
\end{equation}

Which intuitively means that for each $s_i$, there exists a
transition to the (renamed) state $s_{i+1}\cdot \gamma_{i}$.

\subsubsection{Construction of the model}\label{sec:construction}

In this section we show how the \tccp{} Structure that represents the
system behavior is constructed from a labeled specification $S$ in an
automatic way. We present the pseudo-code of the necessary
algorithms for the construction. Moreover, we show the complexity of
such algorithms and explain the process from a theoretical point of
view.

Intuitively, the construction evolves as follows. A process is
composed by a set of clauses and a goal. A \emph{specification} is a
set of clauses. We describe how a specification (or declaration) can
be transformed in a set of \tccp{} Structures.  Actually, for each
different clause we construct a \tccp\ Structure which is labeled
with a unique name.  This name can be used as one of the labels
introduced in the program and is used when a procedure call refers to
such clause.  We consider that the declaration $D_l$ of the form
$\aprocedure{p}{x}_l$ :- $A_l$ is a public information which is always
available. We also assume that each label $l_A$ is associated with the
agent $A$.

The first algorithm that we show is the main procedure
\textsf{construct}($D$) (Figure~\ref{fig:alg-construct}) which, given
a \tccp{} declaration $D_l$ of the form $l_p \aprocedure{p}{x} :-
A_l$, returns a \tccp{} Structure $Q = \langle S,S_0,R,C,T\rangle$
representing the behavior of \textsf{p}.

We define globally a data type called \emph{state} which represents a
state of the \tccp{} Structure. We assume that \emph{store} is a
conjunction of constraints and \emph{label} is a set of labels in
$L$.
\begin{center}
\begin{minipage}{4cm}
\begin{tabbing}
\quad\quad \= \quad\quad \= \kill
state :\\
\> st : store;\\
\> $\ell$[\,] : label;
\end{tabbing}
\end{minipage}
\end{center}

In our pseudo-code, we use the \emph{dot notation} to access to the
components of a state. Moreover, we use the notation $[\,]$ for lists
of elements, thus $\ell[\,]$ is a list of labels.  The $\aleph$ value
is a possible value of a store denoting unsatisfiability.

Finally, we simplify the treatment of functions $C$ and $T$.
Although we do not mention them in the algorithm itself, these
functions correspond to the two components of the state structure
of the algorithm.  We also write $R(n,n')$ to describe that nodes
$n$ and $n'$ are related.

\begin{figure}[ht]
    \centering{ \fbox{
    \begin{minipage}{13cm}
        \begin{tabbing}
            \quad\quad \= \quad\quad \= \quad\quad \= \quad\quad \= \quad\quad \= \quad\quad \= \quad\quad \=
            \quad\quad \= \kill
            $\mathsf{construct}(\mathit{input}\, D: \mathrm{\tccp{}\, declaration},
             \mathit{output}\, \langle S,S_0,R,C,T \rangle: \mathrm{\tccp{}\, Structure})$\\
             \>  s : state;\\
             \>  $S'$, $S'_0$ : set of states;\\
             \>  inf[\,] : store;\\
             \>  lab[\,] : set of labels;\\
             \>  j : int;\\[.2cm]

             \> $S'$ = $\emptyset$; \hspace{10em} // $\emptyset$ denotes the empty set\\
             \> $S'_0$ = $\emptyset$;\\
             \> inf = $\aprocedure{instant}{\true,l_A}$; \hspace{3.8em}// $l_p \aprocedure{p}{x} :- A_l$\\
             \> lab = $\aprocedure{follows}{l_A}$;\\
             \> \textsf{for} $j=1$ \textsf{to} $\aprocedure{sizeof}{\mathrm{inf}}$ \\
             \>\> \textsf{if} inf[j] $<> \aleph$ \textsf{then}\\
             \>\>\> $s =
             \aprocedure{create\_node}{\mathrm{inf[j]},\mathrm{lab[j]}}$;\\
             \>\>\> $C(s) =$ inf[j];\\
             \>\>\> $L(s)$ = lab[j];\\
             \>\>\> $S' = S' \cup \{s\}$;\\
             \> $S_0' = S'$;\\
             \> $\aprocedure{construct\_ag}{S',S_0',R',C',T'}$;\\
             \> $S = S'$; $S_0 = S'_0$; $R = R'$;\\
             \> $C = C'$; $T = T'$;
        \end{tabbing}
    \end{minipage}
    }} \caption{Description of the construction algorithm} \label{fig:alg-construct}
\end{figure}

Roughly speaking, in this algorithm the \tccp{} Structure is
initialized and the set of initial states is created.  Then the
function \textsf{construct\_ag} (Figure~\ref{fig:alg-construct-ag}) is
called. This function iteratively completes the
construction. Functions $\textsf{instant}$ and
$\textsf{follows}$ are two auxiliary procedures used during the
construction of the \tccp{} Structure. We show them below.

Now we show (Figure~\ref{fig:alg-construct-ag}) the
\textsf{construct\_ag} procedure, which uses two more auxiliary
functions: the $\aprocedure{find}{s,S}$ function, which returns a
reference to the state in $S$ which coincides (modulo renaming
of variables) with $s$, and the $\textsf{perm}$ function which,
given two states, returns the necessary renamings which make
them equivalent.

\begin{figure}[ht]
    \centering{ \fbox{
    \begin{minipage}{13cm}
        \begin{tabbing}
            \quad\quad \= \quad\quad \= \quad\quad \= \quad\quad \= \quad\quad \= \quad\quad \= \quad\quad \=
            \quad\quad \= \kill
            $\mathsf{construct\_ag}(\mathit{input/output}\, S[\,] : \mathrm{state};
            \mathit{input}\, S_0[\,]: \mathrm{state}$\\
            \>\>\> \hspace{-.25cm} $\mathit{input/output}\, R : \mathrm{relation}, C, T :\mathrm{function})$\\

             \>  stat1, stat2 : state;\\
             \>  s[\,], acc[\,] : state\\
             \>  inf[\,] : store;\\
             \>  lab[\,] : set of labels;\\
             \>  rn : renaming of variables;\\
             \>  j,k : int;\\[.2cm]

             \> acc = $S$;\\
             \> j = 0;\\
             \> \textsf{while} acc $<> \emptyset$ \textsf{do}\\
             \>\> stat1 = $\aprocedure{select}{\mathrm{acc}}$;\\
             \>\> acc = $\aprocedure{remove}{\mathrm{acc},\mathrm{stat1}}$;\\
             \>\> inf = $\aprocedure{instant}{\mathrm{stat1\ldotp st},\mathrm{stat1}\ldotp\ell}$;\\
             \>\> lab = $\aprocedure{follows}{\mathrm{stat1}\ldotp\ell}$;\\
             \>\> \textsf{for} k=1 \textsf{to} $\aprocedure{sizeof}{\mathrm{inf}}$\\
             \>\>\> \textsf{if} inf[k] $<> \aleph$ \textsf{then}\\
             \>\>\>\> $s[j] =
             \aprocedure{create\_node}{\mathrm{inf[k]},\mathrm{lab[k]}}$;\\
             \>\>\>\> $\mathrm{stat2} = \aprocedure{find}{s[j],S}$;\\
             \>\>\>\> \textsf{if} (stat2) \textsf{then} // there exists an equivalent state\\
             \>\>\>\>\> $rn = \aprocedure{perm}{s[j],stat2}$;\\
             \>\>\>\>\> $R(\mathrm{stat1},rn,\mathrm{stat2})$;\\
             \>\>\>\> \textsf{else}\\
             \>\>\>\>\> $R(\mathrm{stat1},\{\},\mathrm{s[j]})$;\\
             \>\>\>\>\> j = j + 1;\\
             \>\>\>\>\> $S = S\cup$\{s[j]\};\\
             \>\>\>\>\> acc = acc $\cup$ \{s[j]\};\\
             \>\>\>\>\> $C[j] = \mathrm{inf[k]}$;\\
             \>\>\>\>\> $L[j] = \mathrm{lab[k]}$;
        \end{tabbing}
    \end{minipage}
    }} \caption{Description of the construction algorithm for agents} \label{fig:alg-construct-ag}
\end{figure}

Given a label $\mathit{ll}$, \textsf{follows}($ll$) returns the list
which contains the labels associated to the agents that must be
analyzed in the following time instant. Each element of the list
corresponds to a different possible behavior of the system. For
example, in the case of a conditional agent, the initial part of the list
corresponds to the possible behaviors when the guard of the agent is
satisfied, and the final part of the list corresponds to the case when
it is not satisfied. Therefore, if two or more conditional agents are
nested, then all the possible behaviors depending on the first
\emph{then} part will appear before those of the \emph{else} part
in the list. Since \tccp{} restricts the number of nested conditional
agents in a program, we can ensure that this algorithm terminates and
the list of sets of labels is finite.

\begin{figure}[ht]
    \centering{ \fbox{
    \begin{minipage}{10cm}
        \begin{tabbing}
            \quad \= \quad \= \quad \= \quad \= \quad \= \quad \= \quad \=
            \quad \= \kill
            $\mathrm{list\_of\_sets\_of\_stores}\; \mathsf{follows}(ll: \mathrm{label})$\\
            \>\>  $\ell[], \ell_1[], \ell_2[]: \mathrm{set\_of\_labels};$\\
            \>\>  n, i, j : int;\\[.2cm]
            \>\>  \textsf{case} $A$ \textsf{of} \hspace{1cm}// we assume that $A$ is the agent associated with $ll$.\\
            \>\>\>\>  $\astop : \ell[1] = \{\}$;\\
            \>\>\>\>  $\atell{c} : \ell[1] = \{\}$;\\
            \>\>\>\>  $\sum^n_{i=1} \aask{c_i} \rightarrow A_i$ : \textsf{for $j=1$ to $n$}\\
            \>\>\>\>\>\>\> \hspace{2.75cm}$\ell[j] = l_{A_j}$;\\
            \>\>\>\>\>\> \hspace{2.2cm} $\ell[n+1] = \{ll\}$;\\
            \>\>\>\>  $\anow\,c\,\athen\,B_1\,\aelse\,B_2$ : $\ell_1 = \mathsf{follows}(l_{B_1})$;\\
            \>\>\>\>\>\>\> \hspace{2.15cm}$\ell_2 = \mathsf{follows}(l_{B_2})$;\\
            \>\>\>\>\>\>\> \hspace{2.15cm}$\ell = \aprocedure{append}{\ell_1,\ell_2}$;\\
            \>\>\>\>  $B_1\apar B_2 : \ell = \aprocedure{combine}{\aprocedure{follows}{l_{B_1}}, \aprocedure{follows}{l_{B_2}}};$\\
            \>\>\>\>  $\exists x\, B_1 : \ell =  \mathsf{follows}(l_{B_1});$\\
            \>\>\>\>  $\aprocedure{p}{x} : \ell = \{l_p\}$; \hspace{1.15cm} // where $l_p$ represents the label\\
            \>\>\>\> \hspace{3.4cm} // of the \tccp{} Structure constructed for \textsf{p}\\
            \>\>  $\mathsf{end\; case};$\\
            \>\> \textsf{return} $\ell$;
        \end{tabbing}
    \end{minipage}
    }} \caption{Description of the auxiliary algorithm \textsf{follows}($ll$)} \label{fig:alg-follows}
\end{figure}

The \textsf{follows} algorithm uses two additional auxiliary functions,
\textsf{append} and \textsf{combine}, which are functions that
implement operations over lists:
$\aprocedure{append}{\ell_1,\ell_2}$ returns the concatenation of the
two lists $\ell_1$ and $\ell_2$ whereas
$\aprocedure{combine}{\ell_1,\ell_2}$ constructs a new list whose
elements consist of an element of $\ell_1$ and an element of
$\ell_2$. For example, if $\ell_1 = \{\{l_1\},\{l_2\}\}$ and $\ell_2 =
\{\{l_3\}\}$, then the result of $\aprocedure{combine}{\ell_1,\ell_2}$
is the list $\{\{l_1,l_3\},\{l_2,l_3\}\}$.

We can show that the complexity of the algorithm showed in
Figure~\ref{fig:alg-follows} is exponential in the maximum number of
nested agents in the specification. The high complexity is a
theoretical case which does not occur in practice. We think that the
complexity in practical cases should be semi-linear on average.
\begin{lemma}\label{lem:complexity-follows}
  The time complexity for the algorithm \textsf{follows}($A$)
  presented in Figure~\ref{fig:alg-follows} is $O(n*2^m)$ where $m$ is
  the maximum number of nested agents and $n$ is the size of the
   list returned by \textsf{follows}($A$).
\end{lemma}
\begin{proof}First of all, we know that the agent $A$ has a finite number
  of nested agents. Moreover, we can see that the cost of the
  algorithm in the case of Tell and Stop agents is constant since
  \textsf{follows}($A$) = $\{\}$ in such cases. The cost is constant
  also in the case of Procedure Call agents since
  \textsf{follows}($\aprocedure{p}{x}$) returns a single label. For
  the Choice agent, the cost depends on the number of asks contained
  in the agent. Therefore, given the agent $\sum^n_{i=1} \aask{c_i}
  \rightarrow A_i$, the cost will be $n+1$.  In addition, we know that
  the maximum number of nested recursive calls is $2^m$ which
  corresponds to the worst case: when every nested agent is a parallel
  or conditional agent. Note that in these cases, the functions
  \textsf{combine} or \textsf{append} are used. These are indeed the
  expensive operations which we count. We assume that the cost of
  these functions is linear in the size of the resulting list.

Thus, the time complexity of the worst case is $O(n*2^m)$.
\end{proof}

Next we show the second auxiliary function needed during the automatic
construction of the model (see Figure~\ref{fig:alg-instant}).
Given a store and a label,
\textsf{instant}($c,ll$) returns a list of stores which
corresponds to the information which can be computed
instantaneously (\ie{} before the following time instant) by executing
the agents associated with the label $ll$.
\begin{figure}[ht]
    \centering{ \fbox{
    \begin{minipage}{10cm}
        \begin{tabbing}
            \quad \= \quad \= \quad \= \quad \= \quad \= \quad \= \quad \=
            \quad \= \kill
            $\mathrm{list\_of\_stores}\; \mathsf{instant}(\mathit{input}\,\mathrm{st: store}, ll: \mathrm{label})$\\
             \>\>  $s$[], $s_1$[], $s_2$[]: store;\\
             \>\> j: int;\\[.2cm]

             \>\>  \textsf{case} $A$ \textsf{of} \hspace{2cm}// we assume that $ll$ is associated to the agent $A$\\
             \>\>\>\>  $\aabort : s[1] = \true$;\\
             \>\>\>\>  $\atell{c} : s[1] = c$;\\
             \>\>\>\>  $\sum^n_{i=1} \aask{d_i} \rightarrow A_i$ : \textsf{for $j=1$ to $n$}\\
             \>\>\>\>\>\>\> \hspace{2.75cm}s[j] = \{$d_i$\};\\
             \>\>\>\>\>\> \hspace{2.2cm} s[n+1] = $\true$;\\
             \>\>\>\>  $\anow\,d\,\athen\,B_1\,\aelse\,B_2$ : $s_1$ = $\aprocedure{flat}{st, \aprocedure{instant}{st\sqcup d,l_{B_1}}}$;\\
             \>\>\>\>\>\>\>\hspace{2.2cm}$s_2 = \aprocedure{flat}{st,\aprocedure{instant}{\mathsf{not^*}(d)\sqcup st,l_{B_2}}}$;\\
             \>\>\>\>\>\>\>\hspace{2.2cm}$s = \aprocedure{append}{s_1,s_2}$;\\
             \>\>\>\>  $B_1\apar B_2 : s = \aprocedure{combine}{\aprocedure{instant}{st,l_{B_1}},\aprocedure{instant}{st,l_{B_2}}}$\\
            \>\>\>\>  $\exists x\, B_1 : s[1] = \{st[y/x]\} \sqcup \aprocedure{instant}{st,l_{B_1}}$ \hspace{0cm} //where $y$ is a fresh variable\\
            \>\>\>\>  $\aprocedure{p}{x} : s[1] = \true$; \hspace{1cm} // where $\aprocedure{p}{x} :: \{ l_{B_1}\} B_1$ is a\\
            \>\>\>\> \hspace{3.75cm} // clause of the specification\\
            \>\>  $\mathsf{end\; case};$\\
            \>\> \textsf{return} $s$;
        \end{tabbing}
    \end{minipage}
    }} \caption{Description of the auxiliary algorithm \textsf{instant}($st,ll$)} \label{fig:alg-instant}
\end{figure}
In this algorithm we have marked the negation $\mathsf{not(c)}$
with a star to indicate that the semantics of negation is defined
as the non satisfiability of $c$ instead of the satisfiability of
$\neg c$. The $\mathsf{instant}$ procedure uses the auxiliary
function \textsf{flat}(st,$ll$) (Figure~\ref{fig:alg-flat}) which
adds the constraint $st$ to each element of the list $ll$
returning a simple list of stores. If $st$ is inconsistent with
any element of the list, then the value of the element is set to
$\aleph$.

\begin{figure}[ht]
    \centering{ \fbox{
    \begin{minipage}{10cm}
        \begin{tabbing}
            \quad\quad \= \quad\quad \= \quad\quad \= \quad\quad \= \quad\quad \= \quad\quad \= \quad\quad \=
            \quad\quad \= \kill
            $\mathrm{list\_of\_stores}\; \mathsf{flat}(\mathit{input}\,\mathrm{st: store}, ll[]: \mathrm{store})$\\
             \> s[]: store;\\
             \> j: int;\\[.2cm]

             \> \textsf{for} j = 1 \textsf{to} $\aprocedure{sizeof}{ll}$\\
             \>\> \textsf{if} ll[j] $\sqcup$ st = $\mathit{false}$ \textsf{then}\\
             \>\>\> s[j] = $\aleph$;\\
             \>\> \textsf{else} \\
             \>\>\> s[j] = ll[j] $\sqcup$ st;\\
            \> \textsf{return} $s$;
        \end{tabbing}
    \end{minipage}
    }} \caption{Description of the auxiliary algorithm \textsf{flat}(st,$ll$)} \label{fig:alg-flat}
\end{figure}

It is easy to see that the time complexity of \textsf{flat} is linear
on the size of the list.
\begin{lemma}\label{lem:complexity-flat}
  The time complexity for the algorithm \textsf{flat}(c,$ll$)
  presented in Figure~\ref{fig:alg-flat} is $O(n)$ where $n$ is the
  number of elements in the list $ll$.
\end{lemma}
\begin{proof}
  The proof is trivial since we iterate $n$ times over the elements of
  the list.
\end{proof}

The complexity of the algorithm \textsf{instant} showed above is
exponential in the maximum number of nested agents in the
specification. Note that also in this case, this is a theoretical
case which may only occur very rarely in practice.
We think that the complexity
in practical cases should be semi-linear on average.

\begin{lemma}\label{lem:complexity-instant}
  The time complexity for the algorithm \textsf{instant}(st,$A$)
  presented in Figure~\ref{fig:alg-instant} is $O(n*2^m+2n)$ where $m$
  is the maximum number of nested agents and $n$ is the cardinality of
  the list of stores returned by \textsf{instant}(st,$A$).
\end{lemma}
\begin{proof} We know that the agent $A$ has a finite number of nested
  agents. We also know that if the agent is a Stop, Tell or
  Procedure Call agent, then the cost of the function is constant. If
  $A$ is a Choice agent, then we have a linear cost, in particular we
  have $O(n+1)$ since there is an iterative loop over the number $n$
  of guards in the Choice.

  Now let us consider the three remaining cases. For both the
  Conditional and the Parallel agents we have two recursive calls,
  whereas for the Hiding agent we have a single recursive call. We
  assume that the \textsf{combine} and \textsf{append} functions are
  linear in the size of the two lists passed as argument (\ie{} we
  take $O(n)$ where $n$ is the number of elements in the resulting
  list).

  Therefore, we can say that the upper-bound for the global complexity
  of the algorithm is $O(n*2^m+2n)$ where $m$ is the maximum number of
  nested agents.
\end{proof}

Now we can analyze the complexity of the construct algorithm. First of
all, we state the complexity for the \textsf{construct\_ag} function.

\begin{lemma}\label{lem:complexity-construct-ag}
  The time complexity for the algorithm
  \textsf{construct\_ag}($S,S_0,s,R,C,T$) presented in
  Figure~\ref{fig:alg-construct-ag} is $O(c*m*2^m)$ where $m$ is the
  maximum number of nested agents, and $c$ is the number of states in
  the model.
\end{lemma}
\begin{proof}
  By Lemma~\ref{lem:complexity-instant} and
  Lemma~\ref{lem:complexity-follows} we know the complexity of the
  auxiliary functions. Moreover, we know that \textsf{select} and
  \textsf{remove} take linear time and we assume that
  \textsf{create\_node} has constant complexity. We know that the
  \textsf{while} loop will be executed $c$ times, where $c$ is the
  number of different states in the model.

  We can see that each time the loop is executed, we have one
  procedure call to each auxiliary function. Moreover, we have a
  \textsf{for} loop which is executed at most $m+1$ times.  Therefore,
  the cost of the \textsf{for} loop is $O(m)$ and the cost of the
  \textsf{while} loop is $2c*(m*2^m)$. We ensure the finiteness of the
  number of states since we know that there is a finite number of
  combinations of labels and constraints (which appear in the
  specification) modulo renaming.
\end{proof}

\begin{theorem}\label{lem:complexity-construct}
  The time complexity for the algorithm \textsf{construct}($D$)
  presented in Figure~\ref{fig:alg-construct} is $O(c*(2m*2^m))$
  where $m$ is the maximum number of nested agents and $n$ is the
  cardinality of the resulting list.
\end{theorem}

\begin{proof}
  We know the cost of the auxiliary algorithms. Following the
  structure of the algorithm, we can see that there is one call to the
  \textsf{construct\_ag} function. In addition, we have a procedure
  call to the algorithm \textsf{instant} and \textsf{follows}. Then,
  we have to add the cost of such algorithms: $O(2n*2^m+c*(2m*2^m))$.
  We have also a \textsf{for} loop which is executed at most $m$ times.
  Therefore, we obtain the
  global complexity given in this result.
\end{proof}

Let us now explain intuitively the idea of algorithm showed in
Figure~\ref{fig:alg-construct-ag}. Each time an agent is analyzed,
some actions are executed. In the following description we show the
intuitions behind the formal definitions:

\begin{description}
\item[Stop] $S \equiv \astop$. When we find a $\astop$ agent, we add
  no information to the store, insert a self-loop over the new node
  and instantiate the set of labels to the empty set since the
  construction must be concluded.

    \item[Tell] $S \equiv \atell{c}$. The new information $c$ is
    introduced into the store and the label associated to S is
    removed from the labels to be executed.

  \item[Choice] $S \equiv \sum^n_{i=1} \aask{c_i} \rightarrow A_i$.
    This agent leads to a set of corresponding branches in the graph.
    We introduce at most $m+1$ branches with $m\leq n$, one for each
    possible successful ask guard. Note that if a $c_i$ condition is
    not consistent with the store $C(s)$ then the corresponding branch
    will not be generated. For each new node $s'_i$, we define the
    transition $R(s,s'_i\cdot\gamma)$ where $\gamma$ is the
    renaming obtained when new nodes are generated, and we define
    an extra arc $R(s,s_{m+1}\cdot\gamma)$ that corresponds to the
    case when the store does not entail any condition $c_i$ but the
    execution of concurrent agents proceed (if there are no concurrent
    agents or there exist but they cannot proceed, then $s_{m+1}=s$
    thus a self-loop is introduced). Moreover, we do not introduce any
    additional information into the store and the labels are updated.

  \item[Conditional] $S \equiv \anow\, c\, \athen\, A\, \aelse\, B$.
    The construction process in this case follows the same idea as for
    the choice operator: we define two new nodes ($s'_1$ and $s'_2$)
    that correspond to the two possible behaviors. The first branch
    corresponds to the case when the store entails $c$. It is
    added to the store the information that the agent $A$ can
    generate in a single time instant. Also the set of labels is
    updated. The second branch is defined in a similar way.

    \item[Parallel] $S \equiv A\apar B$. When a parallel agent is analyzed, the
    new node generated depends on the execution of the agents $A$ and
    $B$ in the present time instant. This means that the new store is
    defined as the union of the information obtained from the
    execution of $A$ and $B$ (if it is possible to execute them).
    Also the set of labels depends on these two agents.

    \item[Hiding] $S \equiv \exists x\, A$. The behavior of the hiding
    agent is modeled in the graph construction by the introduction of
    the necessary renaming of variables in the store.

  \item[Procedure Call] $S \equiv \aprocedure{p}{X_1,\ldots,X_n}$.
    When a procedure call is reached we finish the process by
    introducing in $s'$ a reference to the initial node of the \tccp{}
    Structure for $\mathsf{p}$. If there are more concurrent agents
    that must be analyzed, then we continue by considering the \tccp{}
    Structure already generated for such clause (with the necessary
    renaming of variables). We link the current node $s$ with a
    simplified copy of this piece of structure. The simplification
    consists in eliminating the branches whose condition is
    inconsistent with the constraints derived by the other (parallel)
    agents.  Thus, the new node $s'$ depends on the execution of the
    other concurrent agents and the body of the clause for
    $\mathsf{p}$.

    If there are two (or more) procedure calls in parallel the
    process is similar and as many nodes as different possible behaviors
    are generated.
\end{description}

\begin{figure}[ht]
    \begin{center}
    \fbox{
       \scalebox{0.4}{\includegraphics{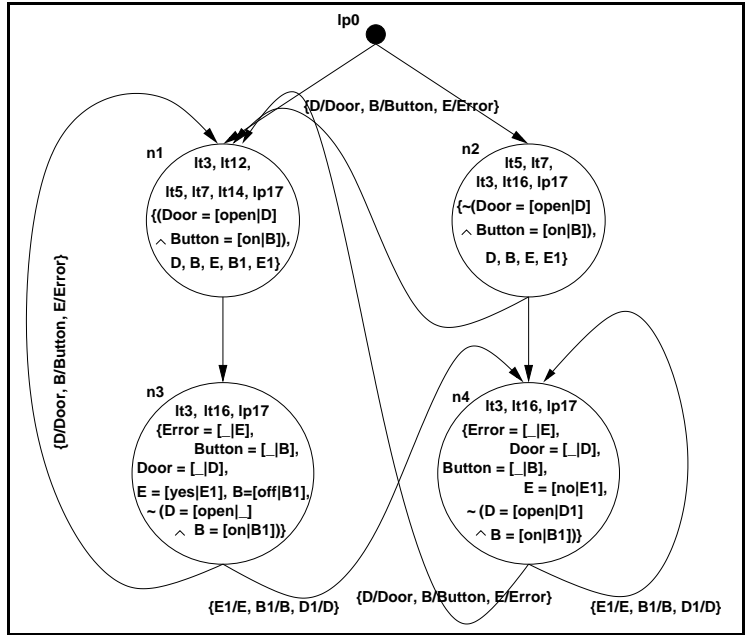}}
    }
    \end{center}
    \caption{Construction of the \protect{\tccp} Structures
    for the example showed in Figure~\ref{fig:tccp-example}} \label{fig:construction}
\end{figure}

In order to illustrate the construction process, in
Figure~\ref{fig:construction} we present the construction of the
\tccp{} Structure for the program in
Figure~\ref{fig:tccp-labelled-example}. Remember that this program
simply detects if the door is open when the microwave works and
in that case turns the system off and emits an error signal.

We can see how, for the first time instant, two nodes corresponding to
the two possible behaviors of the conditional agent have been
generated in the specification ($\mathsf{n1}$ and $\mathsf{n2}$). Now
look at the node $\mathsf{n1}$ where we have that
$L(\mathsf{n1})=\{\mathsf{lt3},\mathsf{lt12},\mathsf{lt5},\mathsf{lt7},
\mathsf{lt14},\mathsf{lp17}\}$.
This means that in order to continue with the graph construction we
have to try to execute the agents associated with such labels.  The
tell agents update the store with the information that an error
combination has been encountered and in the next time instant a
\emph{stop} signal will be present. This is important because when we
try to execute the procedure call associated with $\mathsf{lp}_{17}$,
only one of the two possible branches can be followed.

When we generate new nodes and the corresponding connecting arcs
we should consider formulas which are renamed apart. Note that if
we find a node equal (up to renaming) to another one, a loop will
be formed in the graph and the construction following this branch
will terminate.

\begin{figure}[ht]
    \begin{center}
    \fbox{
       \scalebox{0.35}{\includegraphics{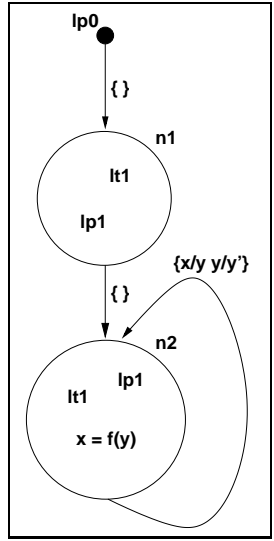}}
    }
    \end{center}
    \caption{Construction of the \protect{\tccp} Structures} \label{fig:example-2}
\end{figure}

Next we show an additional example which may be useful to
understand the construction. Given the program
$$ \aprocedure{p}{x} \text{:-} \exists y (\atell{x = f(y)} || p(y))$$
\noindent the constructed \tccp{} Structure is shown in
Figure~\ref{fig:example-2}. Note that, in each state, we store the new
information added during a single time instant, thus the store of the
program at a time instant $k$ is given by the union of the information
added along the path in the structure, after making $k$ loops. For
instance, after $3$ time instants, the derived information is the
following: $\{x=f(y),y=f(y')\}$.  Roughly speaking, each time we loop
on the second node, a renaming of variables which form the constraints
in the store is performed. Thus, the renaming $\{x/y,y/y'\}$
where $y'$ is a new variable, defines the new constraint $y=f(y')$.
Following the syntax of the program, $x=f(y)$ and $y'=f(y'')$ are
introduced by the tell agent in the first and second time instant
respectively. Note that we show the store after 3 instants of time
since the information produced by a tell agent in a given time instant
(for example, the second), does not appear in the store till the
following time instant. This is due to the fact that tell agents take
one time instant.

\subsubsection{Correctness and Completeness}

In this section we prove the correctness and completeness of the
automatic construction of the model. We first introduce a
function which extracts the information from the
states of the \tccp{} Structure. We define $\mathop{st}$ as the set
of sequences of the form $\{t\mid t = c_1\cdot c_2\cdots
c_n\cdots\}$ where $c_i$ is a finite constraint.

\begin{definition}\label{def:delta}
Given a \tccp{} Structure $Z$ and $s \in tr(Z)$ of the form
$s_0\cdot s_1\cdot\dots$, we define the function
$\delta_s:\mathop{tr(Z)} \rightarrow \mathop{st}$ as follows:

\[    \delta_s(s)=
      \left\{ \begin{array}{ll}  C(s_0) & \text{if $\mathop{s}=s_0$,}\\
                    \delta(s_0)\cdot\delta_s(s') & \text{if $s=s_0 \cdot s'$,}
             \end{array}
      \right.
\]
where $\delta(s_i)$ is defined as
\[
\delta(s_i)= \cup_{0\leq j\leq i} C(s_j)
\]

The extension of $\delta_s$ to sets of sequences is made in the
obvious way.
\end{definition}

The following theorem shows that the defined graph construction is
correct and complete. In other terms it shows that the set of traces
which correspond to the \tccp{} structure $Z$ is the same given by the
operational semantics of the \tccp{} specification $S$.

\begin{theorem}\label{th:tccp-theorem}
    Let $Z$ be the \tccp{} Structure corresponding to the \tccp{}
    specification $S$. Then the construction $Z$ is correct
    and complete since
    \begin{center}
        $\delta_s(tr(Z)) = \cO (S)$
    \end{center}
\end{theorem}
\begin{proof}
  Let us first define an equivalence relation $\sim$ between
  configurations of the operational semantics presented in
  Figure~\ref{fig:tccp-semantics} and graph states. Let
  $\sigma(\Gamma)$ be the store in configuration $\Gamma$,
  then we extend $\sigma$ over sequences of configurations in the
  obvious way. In the graph, stores are `extracted' by using function
  $C$. Then, we say that a configuration $\Gamma$ corresponds to a
  \tccp{} state $s$ if $C(s) \vdash \sigma(\Gamma)$ and
  $\sigma(\Gamma)\vdash C(s)$, and the `active agent' (namely one
  agent immediately reducible given the store in the current
  configuration) in $\Gamma$ corresponds to that selected for
  reduction in $s$; we denote this by $\Gamma \sim s$.

  A trace $t$ of the form $s_0,\ldots,s_i,\ldots$ in a \tccp{}
  Structure $Z$ and a derivation (trace) $\gamma =
  \gamma_0, \ldots,\gamma_i,\ldots$ in the operational semantics of a
  specification $S$ correspond iff $\delta_s(t) = \gamma$, \ie{}
  $\forall i$ $\delta(s_i) \sim \gamma_i$.  We must
  prove that all (the partial) paths in the \tccp{} Structure $Z$
  generated from the specification $S$ have an equivalent trace in the
  operational semantics of $S$ and vice-versa.

  Let us first prove that $\delta_s(tr(Z)) \subseteq \cO (S)$.

  We proceed by induction on the length of the partial trace $n$ in
  $Z$ and on the structure of the agent $A$ selected in step $n$. Note
  that each node in the \tccp{} Structure has a finite number of
  successors, thus we can reason about all of them.

  The basic case for $n=0$ is trivial, since the \tccp{} Structure $Z$
  is based on the same initial state $s_0$ considered in the
  operational semantics.  Let us consider the inductive case, i.e.,
  $n>0$.  Thus, let us consider the trace $s_0,\ldots,s_n $ in $Z$. We
  assume, by inductive hypothesis, that there exists a corresponding
  partial derivation $\gamma_0, \ldots,\gamma_n$ in $\cO (S)$. We now
  prove that, if a further step is made in $Z$ starting from $s_n$, it
  is possible to make a further step starting from $\gamma_n$ in $\cO
  (S)$ and the new states still correspond.  Let $\pi =
  s_0,\ldots,s_n\in\mathop{tr}(Z)$ and let $A$ be the active agent
  selected in $s_n$. We have to consider several cases corresponding
  to the possible structure of $A$.

\begin{description}

\item[Tell] $A = \atell{c}$. Let $C(s_n)=d$ and $T(s_n)=\{l_{tell}\}$.
   Then, we have the trace $\gamma\in\cO(A)$ with $\gamma =
  \gamma_0,\gamma_1,\ldots,\gamma_n$ and $\gamma_n=\langle A,d\rangle$,
  where
   $s_n$ and $\gamma_n$ correspond by inductive hypothesis.
By the definition of the
  construction of the structure and the operational semantics we have
  that $C(s_{n+1})= \{c\sqcup d\}$, $T(s_{n+1})=\{\}$ and
  $\gamma_{n+1}=\langle\emptyset ,c\sqcup d\rangle$ which correspond, thus
  $s_{n+1} \sim \gamma_{n+1}$.

\item[Choice] $A = \sum^m_{i=1}\aask{c_i} \rightarrow A_i$.  Let $C(s_n)=d$ and
  $T(s_n)=\{l_{\mathrm{ask}}\}$. Then, we have the trace
  $\gamma\in\cO(A)$ with $\gamma = \gamma_0,\gamma_1,\ldots, \gamma_{n}$ and
  $\gamma_n=\langle A,d\rangle$. By inductive hypothesis we have that
  $s_n\sim \gamma_n$. By
  definition of the construction of the structure and the operational
  semantics we have two cases: the first case is when there is no
  $c_i$ such that $d\vdash c_i$, then in the construction of the
  \tccp{} Structure there will be a loop, thus the state $s_{n+1}$
  actually is the state $s_n$ whereas in the operational semantics
  there is no possible transition. In this case we just take
  $\gamma_{n+1} = \gamma_n$, and clearly $s_{n+1} \sim \gamma_{n+1}$.

  The second case is when there exists a $c_i$ such that $d\vdash
  c_i$.  This means that we have $C(s_{n+1}) = \{d\}$ and
  $T(s_{n+1})=\{l_{A_i}\}$.  It is clear that by selecting
  $A$ in $\gamma_{n}$, we derive $\gamma_{n+1}$, which
  corresponds to $s_{n+1}$.

\item[Conditional] $A = \anow \,c\, \athen\, A_1\, \aelse\, A_2$.  Let
  $C(s_n)=d$ and $T(s_n)=\{l_{\mathrm{now}}\}$. Then, there exists a
  trace $\gamma\in\cO(A)$ with $\gamma =
  \gamma_0,\gamma_1,\ldots,\gamma_n$ and $\gamma_n=\langle
  A,d\rangle$, such that,
  by inductive hypothesis, $s_n\sim \gamma_n$.
  By definition of the construction of the
  structure and the operational semantics we have two possible
  behaviors: either $d\vdash c$ or $d\not\vdash c$. In the first case,
  $C(s_{n+1}) =
  \{d\sqcup\aprocedure{instant}{d,l_{A_{n+1}}}\}$
  and $T(s_{n+1})=\mathsf{follows}(A_{n+1})$. On the other
  side, we have $\gamma_{n+1}=\langle A_1', d'\rangle$ where $A_1'$ is the
  agent reached by the execution of $A_1$ and $d'$ the new store with
  the information added by the execution of $A_1$. Clearly
  $s_{n+1}$ and $\gamma_{n+1}$ correspond. The case when $d\not\vdash c$ is
  similar, considering $A_{2}$ for reduction.

\item[Parallel] $A = A_1 \apar A_2$.  Let $C(s_n)=d$ and
  $T(s_n)=\{l_{\mathrm{||}}\}$. Then, we have the trace
  $\gamma\in\cO(A)$ with $\gamma = \gamma_0,\gamma_1,\ldots, \gamma_n$ and
  $\gamma_n=\langle A,d\rangle$. By inductive hypothesis
  $s_n\sim \gamma_n$. By
  definition of the construction of the structure and the operational
  semantics we have that $C(s_{n+1}) = \{d\sqcup\aprocedure{instant}{d,l_{A_{n+1}}}\sqcup\aprocedure{instant}{d,l_{A_2}}\}$
  and $T(s_1)=\{\mathsf{follows}(A_1)\cup\mathsf{follows}(A_2)\}$.
  Then, we have $\gamma_{n+1}=\langle A_1'\apar A_2', d'\rangle$
  where $A_1'$ ($A_2'$) is the agent reached by the execution of $A_1$
  ($A_2$) and $d'$ is the new store with the information added by the
  execution of $A_1$ and $A_2$. Hence $s_{n+1} \sim \gamma_{n+1}$.

\item[Exists] $A = \exists x\, A_1$.  Let $C(s_n)=d$ and
  $T(s_n)=\{l_{\mathrm{e}}\}$. Then, we have the trace
  $\gamma\in\cO(A)$ with $\gamma = \gamma_0,\gamma_1,\ldots,\gamma_n$
  and $\gamma_n=\langle A,d\rangle$.  By inductive hypothesis $s_n\sim
  \gamma_n$. We know that $C(s_{n+1}) =
  \{d\sqcup\aprocedure{instant}{d,l_{A_1[y/x]}}\}$. Note that
  $\aprocedure{instant}{d,l_{A_1[y/x]}}$ represents the information
  generated in one time step by the agent $A_1[y/x]$ which is the
  result of the application of the substitution $y/x$ to the agent
  $A_1$ and $T(s_{n+1})=\mathsf{follows}(A_1)$. $y$ is a fresh
  variable, thus the information generated by $A_1$ involving such
  variable will not affect the rest of the system.

  Now, following the operational semantics we derive that $\gamma_{n+1} =
  \langle\exists^{e'}x B,d\sqcup \exists_x e'\rangle$, where
  $\langle A_1,\exists_x d\rangle\rightarrow \langle
  B,e'\rangle$. Thus, we can identify $e'$ with the information
  generated from agent $A_1$, and $s_{n+1}$ and
  $\gamma_{n+1}$ correspond.

\item[Procedure Call] $A = \mathsf{p}(X)$.  Let $\aprocedure{p}{X}:-
  B$ be a clause in the program (in the specification $S$).  Let
  $C(s_n)=d$ and $T(s_n)=\{l_{p}\}$.  By inductive hypothesis, there
  exists the trace $\gamma=\gamma_0,\gamma_1,\ldots,\gamma_{n}
  \in\cO(A)$ and $\gamma_n=\langle A,d\rangle$. We have that $s_{n+1}
  = N$ where $N$ is the first node of the \tccp{} Structure
  constructed for $\aprocedure{p}{X}$. We have that
  $C(s_{n+1})=C(s_n)$ and $T(s_{n+1})=l_B$. By expanding the procedure
  call in the operational semantics we get $\gamma_{n+1}=\langle B,
  d\rangle$, which clearly corresponds to $s_{n+1}$.
\end{description}

Now we have to prove that $\cO (S) \subseteq \delta_s(tr(Z))$.
This is completely analogous to the inclusion that we have proved.
\end{proof}

\section{Specification of the property}\label{sc:the-logic}

In this section we present the logic which we use in our model
checking algorithm. This is a temporal logic which has also the
ability to handle constraints of a given constraint system. In
\cite{BGM01}, the authors presented a temporal logic for reasoning
about \tccp{} programs. In particular, it is an epistemic logic with
two modalities, one representing the \emph{knowledge}
and the other one representing the \emph{belief}. These
two modalities allow us to reason with the input-output behavior of
programs.

Given an atomic proposition $c$ of the underlying constraint system,
$\cK(c)$ and $\cB(c)$ are formulas of
the logic which mean that $c$ is \emph{known} or $c$ is \emph{belief}
respectively. In other words, $\cB(c)$ holds if the process assumes
that the environment provides $c$ whereas $\cK(c)$ holds if the
information $c$ is produced by the process itself.

The syntax of temporal formulas for this logic is shown below (see
\cite{BGM01} for details):
\begin{definition}\label{def:logic-definition}
    Given an underlying constraint system with set of constraints
    $\cC$, formulas of the temporal logic are defined by
    $$\phi ::= \cK(s) \mid \cB(s) \mid \neg\phi \mid \phi \wedge \psi
    \mid \exists x\phi \mid \dnext \phi \mid \phi\, \cU\, \psi$$
\end{definition}

As for classical temporal logics, it is possible to define other logic
operators such as the \emph{always} or \emph{eventually} operators
from the basic ones. For example, if we want to express that a formula
$\phi$ is satisfied at some point in the future, we write that
$\Diamond \phi = \true\,\cU\, \phi$. To express that a formula $\phi$
is always satisfied, we can write that $\Box(\phi) = \neg(\true\,
\cU\, \neg\phi)$. Moreover, as usual we denote by $\phi
\rightarrow \psi$ the formula $\neg\phi\vee (\phi\wedge\psi)$.

A \emph{reaction} is defined as a pair of constraints of the form
$\langle c,d\rangle$ where $c$ represents the input provided by the
environment and $d$ corresponds to the information produced by the
process itself. Moreover, it holds that $d\geq c$ for every reaction,
\ie{} the output always contains the input.

The truth value of temporal formulas is defined with respect to
\emph{reactive sequences}. $\langle
c_1,d_1\rangle\cdots\langle c_n,d_n\rangle\langle d,d\rangle$
denotes a reactive sequence which consists of a sequence of
reactions. Each reaction in the sequence represents a computation
step performed by an agent at time $i$. Intuitively each pair can
be seen as the input-output behavior at time $i$.

Therefore, given a reactive sequence $s$ we can define the truth
values of formulas. The function $\mathsf{first}(s)$ returns the first
reaction of a sequence, \ie{} if $s = \langle
c_1,d_1\rangle\cdots\langle c_n,d_n\rangle\langle d,d\rangle$ then
$\mathsf{first}(s) = \langle c_1,d_1\rangle$.  $\aprocedure{next}{s}$
returns the sequence obtained by removing the first reaction of it,
\ie{} if $s = \langle c_1,d_1\rangle\cdots\langle
c_n,d_n\rangle\langle d,d\rangle$ then $\mathsf{next}(s)=\langle
c_2,d_2\rangle\cdots\langle c_n,d_n\rangle\langle d,d\rangle$.

We say that $\langle c,d \rangle\models \cB(e)$ if $c\vdash e$, \ie{}
the reaction ``believes'' the constraint $e$ if the first component of
the reaction ($c$) entails $e$. Moreover, $\langle c,d \rangle\models
\cK(e)$ if $d\vdash e$, \ie{} the reaction $\langle c,d
\rangle$ ``knows'' the constraint $e$ if its second component entails
$e$.

\newpage
\begin{definition}[by F. de Boer \emph{et al.}]
    Let $s$ be a timed reactive sequence and $\phi$ be a temporal
    formula. Then we define $s\models\phi$ by:
\[
    \begin{array}{ll}
        s\models \cK(c)  &\mathrm{if}\hspace{0.5cm} \mathsf{first}(s)\models \cK(c)\\
        s\models \cB(c) &\mathrm{if}\hspace{0.5cm} \mathsf{first}(s)\models \cB(c)\\
        s\models \neg\phi &\mathrm{if}\hspace{0.5cm} s\not\models\phi\\
        s\models \phi_1 \wedge \phi_2 &\mathrm{if}\hspace{0.5cm} s\models\phi_1 \mathrm{\;and\;} s\models\phi_2\\
        s\models \exists x\phi &\mathrm{if}\hspace{0.5cm} s'\models\phi \text{ for some } s'\; \text{such that}\; \exists_xs = \exists_xs'\\
        s\models \dnext{\phi} &\mathrm{if}\hspace{0.5cm} \mathsf{next}(s)\models \phi\\
        s\models \phi\cU\psi &\mathrm{if}\hspace{0.5cm} \mathrm{for\; some}\; s'\leq s, s'\models \psi\; \mathrm{and\; for\; all\;} s'<s''\leq s, s''\models \phi
    \end{array}
\]
    \noindent where, for a sequence $s = \langle
    c_1,d_1\rangle\cdots\langle c_n,d_n\rangle$, we define the
    existential quantification $\exists_xs = \langle
    \exists_xc_1,\exists_xd_1\rangle\cdots \langle\exists_xc_n,
    \exists_xd_n\rangle$.
\end{definition}

We say that a formula $\phi$ is valid ($\models \phi$) if and only if
for every reactive sequence $s$, $s\models\phi$ holds.  The reader can
see that the modal operators $\cK$ and $\cB$ are monotonic \wrt{} the
entailment relation of the underlying constraint system.

In this work we want to reason about \tccp{} programs. Since the store
of such programs evolves monotonically along the time, the notion of monotonically
increasing reactive sequences is defined: let $s$ be a reactive
sequence of the form $\langle c_1,d_1\rangle\!\!\cdots\!\!\langle
c_{n-1},d_{n-1}\rangle\langle c_n,d_n\rangle$, then we say that $s$ is
monotonically increasing if it satisfies that $c_i\leq d_i$ and
$d_j\leq c_{j+1}$ for each $i\in\{1,\ldots, n\}$ and $j\in\{
1,\ldots, n-1\}$. From now on we consider only monotonically increasing
reactive sequences. In Table~\ref{tab:propiedades} some properties of
the logic operators are shown.

\begin{table*}[ht]
\label{tab:propiedades}
\caption{Logic Operators Properties}
\begin{tabular}{c}
\hline
\hline
$\cB(c) \rightarrow \dbox{\cB(c)}$\\
$\cK(c) \rightarrow \dbox{\cK(c)}$\\
$\cB(c)\rightarrow\cK(c)$\\
$\cK(c)\rightarrow\dnext{\cB(c)}$\\
\hline
\hline
\end{tabular}
\end{table*}

Therefore, whenever a constraint is believed in a specific time
instant, then it will be believed also in all the following time
instants. Moreover, if a given constraint is known at the present time
instant, then it will be known at every time instant in the future.

Finally, we can define a relation between modal operators. In
particular, we say that if a constraint $c$ is believed at a specific
time instant, then it is also known.  Also, if the constraint $c$ is
known at a specific time instant, then it is believed at the following
one.

The logic presented in this section can be seen as a kind of
linear temporal logic. The reader can see that there are no
quantifiers over alternative paths. It is considered that each
instant of time has only one direct successor. If fact, if we
compare this logic with the classical LTL logic (see \cite{CGP99}
for example) we can see that each temporal operator corresponds to
a temporal operator from LTL.

As we have said in the introduction, in model checking we assume a
closed world in the sense that all the agents which can interact with
the system are modeled. For this reason, the output in a time instant
will always coincide with the input in the following time instant,
\ie{} it is not possible that other information different from the one
generated by the model be introduced as an extra-input in any time
instant. This mean that we can work with simple sequences of stores
instead of working with sequences of reactions. We simply eliminate
(or ignore) the second component of each reaction since it coincides
with the fist one of the subsequent reaction.

>From now, when we speak about sequences in the logic, we mean
sequences of the form $s = s_0,s_1,\ldots$ where each $s_i$ is a store
and we omit the modal operator $\mathcal{K}$. The monotonic
properties described above are maintained.

\subsection{Some examples}

Here we illustrate which kind of properties we are able to specify
using this logic. We refer to the program example in
Figure~\ref{fig:tccp-example}. Remember that such example models a
very simplified program which controls the state of the door of a
microwave.

We could check if it is true that when an error is detected, then the
microwave has been turned-off. Actually, the error has occurred
in the previous time instant since the door was open and the microwave
was working, but the program can emit the error signal only in the
following time instant, and at the same time the microwave should be
turned-off.

The following formula represents such property.
\begin{equation}
\begin{array}{ll} \label{formula}
    \neg(\true\; \cU \neg\exists_{\{\mathrm{Error, E, Button, B}\}}(\!\!\!\!\!\! & \mathrm{Error}=[\mathrm{no} \mid E] \vee \\
   &  (\mathrm{Error}=[\mathrm{yes} \mid E] \wedge
    \mathrm{Button}=[\mathrm{off} \mid B])))
\end{array}
\end{equation}

\noindent It could seem that it is a complicate formula but if we
think in terms of the always and eventually operators defined
before, it becomes a very intuitive formula:
$$\Box\exists_{\{\mathrm{Error, E, Button,
    B}\}}(\mathrm{Error}=[\mathrm{no}\mid
E]\vee(\mathrm{Error}=[\mathrm{yes}\mid E]\wedge
\mathrm{Button}=[\mathrm{off}\mid B]))$$

We can also model the property that the door will be eventually
closed:
\begin{equation}
\Diamond\exists_{\{\mathrm{Door, D}\}}(\mathrm{Door}=[\mathrm{close}\mid D])
\end{equation}

Let us now remark the importance of the chosen logic in this work.  We
know that states of the \tccp{} Structure represent only partial
information. Therefore, if we want to check properties directly in the
\tccp{} Structure, then we need a logic able to handle partial
information, as is the case of the logic presented in this section.

If we use any classical logic, we should consider each possible
valuation of the variable values for each \tccp{} state. In that case
we had the same problem as in \cite{FPV00a,FPV00b}, \ie{} we would not
take advantage of the compact representation of the system that
constraints can provide. Finally, the model-checking algorithm would
not be effectively applicable for the state-explosion problem.

\section{The algorithm}\label{sec:the-algorithm}

The third and last phase of the model-checking technique is to
define the algorithm which checks if a given temporal formula is
satisfied by the model. The idea of the algorithm is similar to
that for the classical tableau algorithm for the LTL model
checking problem. The first thing is to construct the
\emph{closure} of the formula $\phi$ that we want to verify. Such
closure is reminiscent of the Fischer-Ladner's one
\cite{FL79}.

Actually, if we intend to prove that the model satisfies the
formula $\phi$, then we construct the closure of the negated
formula ($\neg\phi$). The atomic propositions of the logic are
those of the underlying constraint system. The closure of the
negated formula and the \tccp{} Structure are used to construct a
graph structure (called the \emph{model-checking graph}). This
graph structure consists of nodes of the form $(q,\Phi)$ where $q$
is a state of the \tccp{} Structure and $\Phi$ is a set of
formulas from the closure of $\neg\phi$. The constructed graph
structure allows us to verify if the property is satisfied or not
by the system by using well known graph algorithms. In particular,
we look for a path which starts from an initial state and reaches
a \emph{strongly connected component} (SCC) which satisfies some
properties. If such path exists, then we can say that the property
$\neg\phi$ is satisfied, thus $\phi$ is not satisfied in the model
of the system. In this section we describe this process more in
detail.

The construction of the graph combining the formula and the model
might not terminate. It is for this reason that we use the interval of
time which the user provides to the system. This interval imposes a
time limit. If such time limit is reached, the system aborts the
construction of the graph. The idea is that if this occurs, then we
have obtained an over-approximation of the model, which nevertheless
allows us to make useful verifications over the finite graph
calculated.

\subsection{The closure of the formula}\label{sec:closure}

The closure $CL(\phi)$ of a formula $\phi$ allows us to determine its
truth value. Intuitively, it is the set of sub-formulas that can affect
the truth value. This set is used classically to define tableaux
algorithms where sub-formulas are evaluated as follows: simplest formulas
are evaluated first, then more complex formulas are considered.  Thus,
we can say that the closure of $\phi$ ($CL(\phi)$) is the smallest set
of formulas satisfying the following conditions:
\begin{itemize}
    \item $\phi\in CL(\phi)$,
    \item $\neg\phi_1\in CL(\phi)$ iff $\phi_1\in CL(\phi)$,
    \item if $\phi_1\wedge\phi_2\in CL(\phi)$, then $\phi_1,\phi_2\in CL(\phi)$,
    \item if $\exists x\phi_1\in CL(\phi)$, then $\phi_1\in CL(\phi)$,
    \item if $\dnext{\phi_1}\in CL(\phi)$, then $\phi_1\in CL(\phi)$,
    \item if $\neg\dnext{\phi_1}\in CL(\phi)$, then $\dnext{\neg\phi_1}\in CL(\phi)$,
    \item if $\phi_1\cU\phi_2\in CL(\phi)$, then $\phi_1,\phi_2,\dnext{\phi_1
    \cU\phi_2}\in CL(\phi)$.
\end{itemize}

Note that in the case of $\neg\dnext{\phi_1}$ it is necessary to
introduce the formula $\dnext{\neg\phi_1}$ which cannot be
generated by the other rules.

Now we consider the microwave program example. The formula
(\ref{formula}) for which we calculate the closure is that presented
in the previous section.

\begin{example}
  \medskip For the program showed in Figure~\ref{fig:tccp-example} we
  construct the closure of the formula which we want to
  verify, starting from the negation of Formula
  (\ref{formula}). Note that we assume that $\neg\neg\phi = \phi$. We also
  change in the obvious way the disjunction operator into a
  conjunction:

\begin{equation}
    \true\, \cU\, (\neg(\textrm{Error=[no}
    \mid E]) \wedge
    \neg(\textrm{Error=[yes} \mid E] \wedge
    \textrm{Button=[off} \mid B])) \label{negated-formula}
\end{equation}

Then, we show the closure of the formula. Note that the size of the
set of formulas in the closure increases polynomially with the
size of the formula (meaning the number of operators in the
formula).\\[1ex]

\hspace{-1em}$
\begin{array}{l}
CL(\chi) = \{ \true\,\cU\,(\neg(\mathrm{Error}=[\mathrm{no}\mid
E])\wedge
\neg(\mathrm{Error}=[\mathrm{yes}\mid E]\wedge \mathrm{Button}=[\mathrm{off}\mid B])),\\
\hspace{2em} \true,\\
\hspace{2em} \false,\\
\hspace{2em} \neg(\mathrm{Error}=[\mathrm{no}\mid E]) \wedge \neg(
\textrm{Error}=[\mathrm{yes}\mid E] \wedge
\mathrm{Button}=[\mathrm{off}\mid B]),\\
\hspace{2em} \neg(\mathrm{Error}=[\mathrm{no}\mid E]),\\
\hspace{2em} \neg(\textrm{Error}=[\mathrm{yes}\mid E] \wedge
\mathrm{Button}=[\mathrm{off}\mid B]),\\
\hspace{2em} \neg(\neg(\mathrm{Error}=[\mathrm{no}\mid E]) \wedge
\neg(\textrm{Error}=[\mathrm{yes}\mid E] \wedge
\mathrm{Button}=[\mathrm{off}\mid B])),\\
\hspace{2em} \mathrm{Error}=[\mathrm{no}\mid E],\\
\hspace{2em} \mathrm{Error}=[\mathrm{yes}\mid E] \wedge
\mathrm{Button}=[\mathrm{off}\mid B],\\
\hspace{2em} \mathrm{Error}=[\mathrm{yes}\mid E],\\
\hspace{2em} \mathrm{Button}=[\mathrm{off}\mid B],\\
\hspace{2em} \neg(\mathrm{Error}=[\mathrm{yes}\mid E]),\\
\hspace{2em} \neg(\mathrm{Button}=[\mathrm{off}\mid B]),\\
\protect{\hspace{2em}} \dnext{\true\,\cU\,(\neg(\mathrm{Error}=[\mathrm{no}\mid
E])\wedge
\neg(\mathrm{Error}=[\mathrm{yes}\mid E]\wedge \mathrm{Button}=[\mathrm{off}\mid B]))},\\
\hspace{2em} \neg(\dnext{\true\,\cU\,(\neg(\mathrm{Error}=[\mathrm{no}\mid
E])\wedge
\neg(\mathrm{Error}=[\mathrm{yes}\mid E]\wedge \mathrm{Button}=[\mathrm{off}\mid B]))}),\\
\protect{\hspace{2em}} \dnext{\neg(\true\,\cU\,(\neg(\mathrm{Error}=[\mathrm{no}\mid
E])\wedge
\neg(\mathrm{Error}=[\mathrm{yes}\mid E]\wedge \mathrm{Button}=[\mathrm{off}\mid B])))},\\
\hspace{2em} \neg(\true\,\cU\,(\neg(\mathrm{Error}=[\mathrm{no}\mid E])\wedge
\neg(\mathrm{Error}=[\mathrm{yes}\mid E]\wedge
\mathrm{Button}=[\mathrm{off}\mid B])))\\
\hspace{2em} \}
\end{array}
$
\end{example}

\subsection{The model-checking graph}

Given a formula $\phi$ of the logic described in
Section~\ref{sc:the-logic}, and the \tccp{} Structure $Z$ constructed
from the specification, the graph $G(\phi,Z)$ is defined as follows
\begin{definition}[Model-Checking Graph]\label{def:graph-G}
    Let $\phi$ be a formula, $CL(\phi)$ the closure of $\phi$ as defined in Section~\ref{sec:closure} and $Z$
    the \tccp{} Structure constructed following the algorithm
    described in Section~\ref{sec:construction}. A node $n$ of the model-checking graph
    is formed by a pair of the form $(s_n, \cQ_n)$ where $s_n$ is
    a state of $Z$ and $\cQ_n$ is a subset of $CL(\phi)$ and the
    atomic propositions such that the following conditions are
    satisfied:
    \begin{itemize}
        \item for each atomic proposition $p$, $\cK(p)\in \cQ_n$ iff $p\in
        C(s_n)$,
        \item for every $\exists x\phi_1\in CL(\phi),
        \exists x\phi_1\in\cQ_n$ iff $\exists_x\phi_1\in C(s_n)$,
        \item for every $\phi_1\in CL(\phi), \phi_1\in \cQ_n$ iff
        $\neg\phi_1\not\in \cQ_n$,
        \item for every $\phi_1\wedge\phi_2\in CL(\phi),
        \phi_1\wedge\phi_2\in\cQ_n$ iff $\phi_1\in\cQ_n$ and
        $\phi_2\in\cQ_n$,
        \item for every $\neg\dnext\phi_1\in CL(\phi),
        \neg\dnext\phi_1\in\cQ_n$ iff $\dnext\neg\phi_1\in\cQ_n$,
        \item for every $\phi_1\,\cU\,\phi_2\in CL(\phi),
        \phi_1\,\cU\,\phi_2\in\cQ_n$ iff $\phi_2\in\cQ_n$ or
        $\phi_1,\dnext{
        \phi_1\,\cU\,\phi_2}\in\cQ_n$.
    \end{itemize}

    An edge in the graph is defined as follows: there will be an
    edge from one node $n_1=(s_1,Q_1)$ to another node $n_2=(s_2,Q_2)$
    iff  there is an arc from the node $s_1$ to the node $s_2$ in the
    \tccp{} Structure and for every formula $\dnext\phi_1\in
    CL(\phi)$, $\dnext\phi_1\in Q_1$ iff $\phi_1\in Q_2$.
\end{definition}

Note that, in the definition above, when we take into
consideration the set of arcs of the \tccp{} Structure (when
analyzing the formulas containing the next operator), we also
consider the renaming that may label these arcs.

Intuitively, for each node of the model-checking graph, in $\cQ$ we
have the largest consistent set of formulas that is also consistent
with the labelling function (the function $C$) of the \tccp{}
Structure.  Moreover, two nodes of the graph are related if the
temporal formulas in their $\cQ$ sets are consistent.

For each node $s_i$ of the \tccp{} Structure many nodes are
generated in the model-checking graph. All these nodes have as
first component the state $s_i$ and the second component consists
of the different consistent sets of formulas derived from $C(s_i)$
and the closure of the formula.

Next we show an example to illustrate how the nodes of the
model-checking graph are constructed. We construct the graph for the
negation of the property since we intend to prove that there is no
computation of the system which satisfies the negated property. This
is equivalent to prove that the property is satisfied for all the
computations.

\begin{example}\label{ex:G-graph}
\medskip
In this example we show some nodes of the graph which would result
from our program example. We take the \tccp\ Structure shown in
Figure~\ref{fig:construction} and the closure set of the formula
showed in the previous section.

 Here we show two of the nodes generated for
$s_1$ and one of the nodes generated for $s_2$.

\hspace{-1em}$
\begin{array}{rl}
n_1=&(s_1,Q_1)\; \mathrm{where}\\
Q_1= &\{\\
&\mathrm{Door} =[\mathrm{open}\mid D]\wedge
\mathrm{Button} = [\mathrm{on}\mid B],\\
&\true,\;\mathrm{Error} = [\mathrm{no}\mid E],\\
&\neg(\mathrm{Button}=[\mathrm{off}\mid B]),\\
&\neg(\mathrm{Error} = [\mathrm{yes}\mid E]\wedge
\mathrm{Button}=[\mathrm{off}\mid B]),\\
&\neg(\neg(\mathrm{Error}=[\mathrm{no}\mid E])\wedge
\neg(\mathrm{Error}=[\mathrm{yes}\mid E]\wedge
\mathrm{Button}=[\mathrm{off}\mid B])),\\
&\dnext{\true\;\cU\;(\neg(\mathrm{Error}=[\mathrm{no}\mid
E])\wedge \neg(\mathrm{Error}=[\mathrm{yes}\mid E]\wedge
\mathrm{Button}=[\mathrm{off}\mid B]))},\\
&\true\;\cU\;(\neg(\mathrm{Error}=[\mathrm{no}\mid E])\wedge
\neg(\mathrm{Error}=[\mathrm{yes}\mid E]\wedge
\mathrm{Button}=[\mathrm{off}\mid B]))\\
&\}
\end{array}
$

\hspace{-1em}$
\begin{array}{rl}
n_2=&(s_1,Q_2)\; \mathrm{where}\\
Q_2=&\{\\
&\mathrm{Door} =[\mathrm{open}\mid D]\wedge
\mathrm{Button} =
[\mathrm{on}\mid B],\\
&\true,\;\mathrm{Error} = [\mathrm{yes}\mid E],\\
&\neg(\mathrm{Button}=[\mathrm{off}\mid B]),\\
&\neg(\mathrm{Error} = [\mathrm{yes}\mid E]\wedge
\mathrm{Button}=[\mathrm{off}\mid B]),\\
&\neg(\mathrm{Error}=[\mathrm{no}\mid E])\wedge
\neg(\mathrm{Error}=[\mathrm{yes}\mid E]\wedge
\mathrm{Button}=[\mathrm{off}\mid B]),\\
&\true\;\cU\;(\neg(\mathrm{Error}=[\mathrm{no}\mid E])\wedge
\neg(\mathrm{Error}=[\mathrm{yes}\mid E]\wedge
\mathrm{Button}=[\mathrm{off}\mid B])),\\
&\dnext{\true\;\cU\;(\neg(\mathrm{Error}=[\mathrm{no}\mid
E])\wedge \neg(\mathrm{Error}=[\mathrm{yes}\mid E]\wedge
\mathrm{Button}=[\mathrm{off}\mid B]))}\\
&\}
\end{array}
$

\hspace{-1em}$
\begin{array}{rl}
n_3=&(s_2,Q_3)\;\mathrm{where}\\
Q_3=&\{\\
&\mathrm{Error}=[\mathrm{yes}\mid E],\;
\mathrm{Button}=[\mathrm{off}\mid B],\\
& \neg(\mathrm{Door}=[\mathrm{open}\mid D]\wedge
\mathrm{Button}=[\mathrm{on}\mid B]),\\
&\true,\\
&\mathrm{Error}=[\mathrm{yes}\mid E]\wedge
\mathrm{Button}=[\mathrm{off}\mid B],\\
&\neg(\neg(\mathrm{Error}=[\mathrm{no}\mid E])\wedge
\neg(\mathrm{Error}=[\mathrm{yes}\mid E]\wedge
\mathrm{Button}=[\mathrm{off}\mid B])),\\
&\dnext{\true\;\cU\;(\neg(\mathrm{Error}=[\mathrm{no}\mid
E])\wedge \neg(\mathrm{Error}=[\mathrm{yes}\mid E]\wedge
\mathrm{Button}=[\mathrm{off}\mid B]))},\\
&\true\;\cU\;(\neg(\mathrm{Error}=[\mathrm{no}\mid E])\wedge
\neg(\mathrm{Error}=[\mathrm{yes}\mid E]\wedge
\mathrm{Button}=[\mathrm{off}\mid B]))\\
\}
\end{array}
$

Then, following the definition of the model-checking graph, we can
define an arc from $n_2$ to $n_3$ since for each formula of the
form $\dnext{\phi}$ in the closure, if it is in $Q_2$ then $\phi$
is in $Q_3$.
\begin{figure}[ht]
    \begin{center}
    \fbox{
       \scalebox{0.4}{\includegraphics{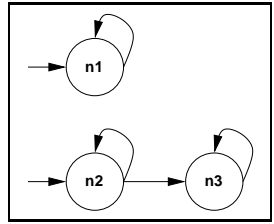}}
    }
    \end{center}
    \caption{A part of the model-checking graph for the
    \protect{\tccp} Structure showed in Figure~\ref{fig:construction}
    and the Formula (\ref{formula})} \label{fig:modelcheckinggraph}
\end{figure}
\end{example}

In this example, a brief time interval is sufficient to build the
complete graph without approximation. During the construction, we can
annotate how many steps are needed to reach each node from a root
note, which determines the current instant of time. If such instant of
time is equal to the time limit, then the construction is concluded
and the graph obtained since that moment is given as output of the
algorithm.

\subsection{The searching algorithm}

It is well known that in order to prove that a
property is satisfied, it
is possible to prove that there is no path satisfying
the negation of the property. Thus, for verifying the formula
$\phi$, we construct the model-checking graph using the negation of
$\phi$ and the model of the system. Then we look for a sequence such
that starting from the initial node of the graph, it reaches a
\emph{self-fulfilling strongly connected component} (SCC).  Let us now
give the formal definitions of SCC and self-fulfilling SCC.

\begin{definition}[Strongly Connected Component]\label{def:SCC}
Given a graph $G$, we define a Strongly Connected Component (SCC)
$C$ as a \emph{maximal} subgraph of $G$ such that every
node in $C$ is reachable from every other node in $C$ along a
directed path entirely contained within $C$.

We say that $C$ is \emph{nontrivial} iff
either it has more than one node or it contains one node with a
self-loop.
\end{definition}

Then we can define a kind of strongly connected component.
Actually, we will search for SCC satisfying the following
properties in our model-checking algorithm.

\begin{definition}[Self-fulfilling SCC]\label{def:SSCC} Given a
model-checking graph $G$, a self-fulfilling strongly connected
component $C$ is defined as a nontrivial strongly connected
component in $G$ which satisfies that for every node $n$ in $C$
and for every $\phi_1\,\cU\,\phi_2\in Q_n$, there exists a node
$m$ in $C$ such that $\phi_2\in Q_m$, and vice-versa.
\end{definition}

Now, let $G$ be the model-checking graph generated following the steps
described in Definition~\ref{def:graph-G}. We say that a sequence is
an \emph{eventually sequence} if it is an infinite path in $G$ such
that if there exists a node $n$ in the path with
$\phi_1\,\cU\,\phi_2\in\cQ_n$, then there exists another node $n'$ in
the same path reachable from $n$ along the path, such that
$\phi_2\in\cQ_{n'}$.

Moreover, we can prove the following result, which says that if we
find a self-fulfilling strongly connected component in the
corresponding model-checking graph, then the property represented by
the formula is satisfied by the system. Our problem will be to prove
that such self-fulfilling SCC does not exist\footnote{Note that the
  result assumes that the construction of the graph has terminated
  before reaching the time limit provided by the user.}.
\begin{theorem}
    Let $\phi$ be a formula, $Z$ a \tccp{} Structure and
    $G(\phi,Z)$ the corresponding model-checking graph. If there
    exists a path in $G$, which satisfies a formula
    $\phi$, from an initial node
    to a self-fulfilling strongly connected component, then
    the model $Z$ satisfies the formula $\phi$.
\end{theorem}
\begin{proof}
  In order to prove this theorem we prove instead an equivalent
  result. We prove that if there exists an eventually sequence
  starting at an initial node $n=(s,\cQ_n)$ such that the formula
  $\phi$ is in $\cQ_n$, then the model satisfies the formula $\phi$.
  This result is equivalent to the statement of the theorem since
  classical results \cite{CGP99,MP95} show that there exists an
  eventually sequence starting at a node $n=(s,\cQ_n)$ if and only if
  there is a path in $G(\phi,Z)$ from $n$ to a self-fulfilling SCC. We
  show that we can extend this result directly to our framework.

  Assume that we have an eventually sequence $n_1,n_2,\ldots$ where
  $n_1=(s_1,\cQ_{n_1})$, $n_2=(s_2,\cQ_{n_2})$, etc., starting with
  $n_1=n$. This eventually sequence starts at node $n_1$ with
  $\phi\in\cQ_{n_1}$. By definition, $\pi=s_1,s_2,\ldots$ is a path in
  the model $Z$ starting at $s = s_1$. We want to show that
  $\pi\models\phi$. We will prove a stronger result: for every formula
  $\psi$ in the closure of the formula $\phi$ ($\psi\in CL(\phi)$) and
  every $i\geq 0,\pi^i\models\psi$ iff $\psi\in\cQ_{n_i}$. We follow
  the classical notations and by $\pi^i$ with $i\geq 0$ we mean the
  suffix of the path $\pi$ starting from the $i$-th component: $\pi^i
  = s_i,s_{i+1},\ldots$.  The proof proceeds by structural induction
  over the sub-formulas.  There will be six cases corresponding to the
  six considered operators of the logic.
\begin{enumerate}
\item If $\psi$ is an atomic formula, then by
  Definition~\ref{def:graph-G} of node $n_i$, $\psi\in\cQ_{n_i}$ iff
  $\psi\in C(s_i)$.
    \item if $\psi=\exists x\chi$ then $\pi^i\models\psi$ iff
    $\psi\in C(s_i)$.
    \item If $\psi = \neg\chi$ then $\pi^i\models \psi$ iff
    $\pi^i\not\models\chi$. By the inductive hypothesis, this holds
    iff $\chi\not\in\cQ_{n_i}$. By Definition~\ref{def:graph-G}, this
    guarantees that $\psi\in\cQ_{n_i}$.
    \item If $\psi = \chi_1\wedge\chi_2$ then $\pi^i\models\psi$ iff
    $\pi^i\models\chi_1$ and $\pi^i\models\chi_2$. By the inductive
    hypothesis, this holds iff $\chi_1\in\cQ_{n_i}$ and
    $\chi_2\in\cQ_{n_i}$. By Definition~\ref{def:graph-G} this is true iff
    $\psi\in\cQ_{n_i}$.
    \item if $\psi = \dnext\chi$ then $\pi^i\models\psi$ iff
    $\pi^{i+1}\models\chi$. By the inductive hypothesis this holds iff
    $\chi\in\cQ_{n_{i+1}}$. Since
    $((s_i,\cQ_{n_i}),(s_{i+1},\cQ_{n_{i+1}}))\in R$, the above holds
    iff $\dnext\chi\in\cQ_{n_i}$.
    \item if $\psi = \chi_1\,\cU\,\chi_2$ then by definition of an
    eventually sequence, there is some $j\geq i$ such that
    $\chi_2\in\cQ_{n_j}$. Since $\psi\in\cQ_{n_i}$, the definition of
    a node implies that if $\chi_2\not\in\cQ_{n_i}$, then
    $\chi_1\in\cQ_{n_i}$ and $\dnext\psi\in\cQ_{n_i}$. In this case,
    the definition of the transition relation of $G$ implies that
    $\psi\in\cQ_{n_{i+1}}$. It follows that for every $i\leq k < j$,
    $\chi_1\in\cQ_{n_k}$. By the inductive hypothesis,
    $\pi^j\models\chi_2$ and for every $i\leq k<j$,
    $\pi^k\models\chi_1$. Hence $\pi^i\models \psi$.

    Since $\pi^i\models\psi$, then there exists $j\geq i$ such that
    $\pi^j\models\chi_2$ and for all $i\leq k < j$,
    $\pi^k\models\chi_1$. We take the minimum $j$. By the inductive
    hypothesis, $\chi_2\in\cQ_{n_j}$ and for every $i \leq k< j$,
    $\chi_1\in\cQ_{n_k}$. Suppose $\psi\not\in\cQ_{n_i}$. Since
    $\chi_1\in\cQ_{n_i}$, by Definition~\ref{def:graph-G}
    $\dnext\psi\not\in\cQ_{n_i}$, which implies that
    $\dnext\neg\psi\in\cQ_{n_i}$. Now by definition of the transition
    relation of $G$, $\neg\psi\in\cQ_{n_{i+1}}$, and hence
    $\psi\not\in\cQ_{n_{i+1}}$. Continuing the argument inductively,
    we would eventually find $\psi\not\in\cQ_{n_k}$, which is a
    contradiction since $\chi_2\in\cQ_{n_j}$.
\end{enumerate}
This proves that if we have an eventually sequence, the model
satisfies the formula $\phi$. Now we have the classical result that
can be applied to the graph $G$. If we look for an eventually sequence,
we can instead look for a path from the initial node $n$ to a
self-fulfilling SCC. There are algorithms that implement this search
with a complexity linear in the size of the graph and exponential in
the size of the formula.
\end{proof}

For the complexity of the algorithm, we can see that
the method is quite inefficient since it is based on the tableau
algorithm for LTL. Note that such algorithm is PSPACE-complete.
The important thing is the fact that we are dealing with a
programming language and we can handle constraints as a powerful
way to represent systems. Moreover, we obtain a similar
complexity to the classical approach since we use a logic which
is able to handle \tccp{} states. If we had used a classical logic, the
complexity would have increased too much since it would be necessary to unfold
the states of the graph structures in order to consider all the
possible valuations of variables which could satisfy a given
constraint.

\section{Related Works}\label{related-work}

We can find in the literature some related works which use the notion
of constraint in order to solve the automatic verification problem for
infinite-state systems. For example, in \cite{DP01} and \cite{DP99},
the authors introduce a methodology to translate concurrent systems
into CLP programs and verify safety and liveness properties over such
CLP programs. \cite{EM97} introduces a semi-decision algorithm that
uses constraint programming in order to verify 1-safe Petri nets.
Actually, while in \cite{DP01,DP99}, constraints are used as an
abstract representation of sets of system states, in \cite{EM97}
constraint programming is used for solving linear constraints in the
implementation of the algorithm.

Constraints are useful for different purposes in
software verification. They can be used in the checking
algorithms as is done in \cite{EM97}; they can be used to model the
problem as Delzanno and Podelsky do; and  they can also be integrated
into the specification language,
that is used to model the system, as we do.

Regarding the systems that our approach is able to verify, we have
seen that there are basically two main cases. The first case is when
we are able to verify a system without the limitation on the
time interval and the second case is when the time limit is reached.
The first case corresponds to systems whose infinite nature comes from
the fact that they use variables with an infinite domain. These
systems are somehow similar to the ones that can be verified in
\cite{DP01} for the properties of safety. In the second case
we consider a large class of systems by using the time interval
``approximation''.  If we reach the limit of time imposed by the user
(obviously, if the user provides a too short time interval, then some
systems of the first class end up in this second category) then we
must stop the construction of the graph $G$ at that point. Thus,
we can verify the system, but we must consider that it is an
approximation of the original system.

We note that there are some
limitations in the \tccp{} language since, for example, \tccp{} is not
able to model strong preemption while \cite{DP01} considers a
language which can express this behavior.

In the last years many different extensions over time have been
presented in the literature. There are approaches which extend the
\cc\ paradigm with a notion of \emph{discrete} time (\tccp{}, \tcc{}
\cite{SJG94a} or \ntcc{} \cite{Val02}) and there is also an extension
of the model with a notion of \emph{continuous} notion of time
(\emph{hybrid} \cc{} language \cite{GJS98}). Regarding \ntcc{}, in
\cite{Val03}, the author presented some decidability results with
respect to such language. Those results show that it is possible to apply
model checking to \ntcc{} but no algorithm nor complexity studies are
presented.

In \cite{FPV00a,FPV00b} a method to construct a
structure was presented
as a first step towards the definition of a model-checking
technique for \tcc{}. Nevertheless, the structure defined in
\cite{FPV00a,FPV00b} to model \tcc{} programs was quite different from
the structure defined in this work. Actually, in those works
the modeling phase was defined in detail, giving only a brief description
of the specification and the algorithmic phase.

The \tcc{} structure had two kind of transitions: the \emph{timed
  transitions} and the \emph{normal transitions}. The set of states of
the \tcc{} Structure were defined in a way as similar to the \tccp{}
Structure and could also be seen as sets of classical states for a
Kripke Structure.  However, also in this case, classical model
checking algorithms cannot be applied to \tcc{} Structures. First of
all because \tcc{} Structures have two kind of transitions, and
secondly because the algorithms cannot handle the notion of state of
the graph structure.  Note that in the \tccp{} approach we have only
one kind of transition relation, thus we have only one problem: how to
handle states.

Another main difference between the \tcc{} and the \tccp{} Structure
lies in the interpretation of branching points.  Branching points in
\tcc{} Structures are  due to the interleaving nature of the model.
The \emph{normal} transitions are instantaneous in the sense that
they do
not cause time steps.  The branching points of the \tccp{}
Structure due to conditional agents can be viewed as the branching
points which could appear in the quiescence points of the \tcc{}
Structure, \ie{} when passing from one time instant to the following
one. However, branching points of the \tccp{} Structure due to Choice
agents cannot be identified with anything in the \tcc{} Structure
since the \tcc{} model is deterministic.

In \cite{FPV00a,FPV00b}
the idea was to transform the \tcc{} Structure into a Kripke
structure, and hence the problem at this point was the huge number of
states of the transformed structure. Essentially, we lost the
possibility to take advantage of the compact representation that
the notion of constraint provides.

In the \tccp{} approach it is not necessary to eliminate the kind of
transitions (since there is only one type). More important is the fact
that it is not necessary to unfold the possible values of variables in
order to define a model-checking method. Actually, we use a temporal
logic which is able to handle the \tccp{} states.

In \cite{FPV01}  a first approach to the
problem of verification of \hcc{}, which is similar to the problem for
\tccp{} was presented.  The idea was the essentially similar, \ie{} to
define a structure able to
represent the system behavior and to check properties over such
structure.  However, we just constructed the basic model
which was transformed into a linear time automaton which could be given as
input to a classical model checker such as \textsc{HyTech}.

\section{Conclusions}\label{conclusions}

In this work we have introduced a method that allows us to check
properties from a temporal logic over reactive systems that are
specified in the Temporal Concurrent Constraint Language defined in
\cite{BGM99}. We have seen that we can adapt the classical method of
LTL model checking to the logic presented in \cite{BGM01} and the {\tt
  tccp} Structure defined in this paper which models the system
behavior. We have described a method that can handle generic programs
written in {\tt tccp}, which means that we are not restricting
ourselves to finite-state systems. By using {\tt tccp} we can define
infinite-state systems that can be handled by the logic which we have
used. This epistemic logic allows us to work with constraints.
Constraints can be seen as a compact representation of (possibly
infinite) many states. In a previous work \cite{FPV00a,FPV00b} the
authors have defined a structure which can help to verify a different
class of reactive systems specified using another language from the
{\tt ccp} framework. \cite{FPV00a,FPV00b} defined a kind of
structure that may seem similar to the {\tt tccp} Structure but it is
essentially different: the nodes and the arcs of the
graph structure are interpreted in a different manner. Furthermore
\cite{FPV00a,FPV00b} do not define any model-checking algorithm,
rather they only concentrate on the modeling phase. We have proved
that our verification method is correct and have illustrated
how it works.

We plan to make a prototypical implementation of our system and
test it on a set of benchmarks, such as protocol verification and
verification of properties of concurrent systems like safety or
liveness properties.

We also want to study how our method can be optimized in order to
improve its efficiency. It is well known that this kind of
classical model-checking algorithm is exponential in the size of
the formula. Hence as future work we want to extend to
our framework some efficient
model-checking algorithms, such as symbolic model checking, for
avoiding a complete construction of the graph.

\end{document}